\documentclass[a4paper,11pt]{article}
\pdfoutput=1 

\usepackage{jcappub} 

\usepackage[T1]{fontenc} 
\usepackage{wrapfig}
\usepackage{verbatim}
\usepackage{subcaption}
\usepackage{graphicx}
\usepackage[normalem]{ulem}
\usepackage{xcolor}

\title{\boldmath Particle content of very inclined air showers for radio signal modeling}

\newcommand{\m}{\color{orange}}

\author[a,b,1]{Marion Guelfand,\note{Corresponding author.}}
\author[b,2]{Simon Chiche,}
\author[b,c, 2]{Kumiko Kotera,\note{Also at Some University.}}
\author[b,d,2]{Simon Prunet}
\author[e]{Tanguy Pierog}

\affiliation[a]{Sorbonne Université, CNRS, Laboratoire de Physique Nucléaire et des Hautes Energies (LPNHE),4 Pl. Jussieu, 75005 Paris, France}
\affiliation[b]{Sorbonne Université, CNRS, Institut d'Astrophysique de Paris, 98 bis bd Arago, 75014, France}
\affiliation[c]{Vrije Universiteit Brussel, Physics Department, Pleinlaan 2, 1050 Brussels, Belgium}
\affiliation[d]{Université Côte d’Azur, Observatoire de la Côte d’Azur, CNRS, Laboratoire Lagrange, \\ Bd de l’Observatoire, CS 34229, 06304 Nice cedex 4, France}
\affiliation[e]{Insitute for Astroparticle Physics, Karlsruhe Institute of Technology (KIT),
Hermann-von-Helmholtz-Platz 1, 76344 Eggenstein-Leopoldshafen, Germany}

\emailAdd{marion.guelfand@lpnhe.in2p3.fr}
\emailAdd{simon.chiche@iap.fr}
\emailAdd{kumiko. kotera@iap.fr}
\emailAdd{simon.prunet@oca.eu}
\emailAdd{tanguy.pierog@kit.edu}

\abstract{The reconstruction of very inclined air showers is a new challenge for next-generation radio experiments such as the AugerPrime radio upgrade, BEACON, and GRAND, which focus on the detection of ultra high energy particles. To tackle this, we study the electromagnetic particle content of very inclined air showers, which has scarcely been studied so far. 
Using the simulation tools CORSIKA and CoREAS, and analytical modeling, we explore the energy range of the particles that contribute most to the radio emission, quantify their lateral extent, and estimate the atmospheric depth at which the radio emission is strongest. We find that the spatial distribution of the electromagnetic component in very inclined air showers has characteristic features that could lead to clear signatures in the radio signal, and hence impact the reconstruction strategies of next-generation radio-detection experiments.}

\begin{document}
\maketitle
\flushbottom

\section{Introduction}
\label{sec:intro}

Recently, several radio experiments (AERA, CODALEMA, TREND, LOFAR) \cite{AERA, Schr_der_2017} have demonstrated that the information contained in the radio signal of vertical air showers allows one to perform an accurate reconstruction of the primary particle parameters (nature, arrival direction, energy) \cite{Nelles_2015}. The next generation of extended radio arrays (GRAND, BEACON, AugerPrime radio upgrade) \cite{GRAND_paper, beacon, AugerPrime} focuses on very inclined extensive air showers (EAS) that leave large radio footprints on the ground. Thanks to this specific feature, one can deploy detectors over gigantic arrays with a sparse density to provide large detector effective area at affordable costs.

The study of the radio signal produced by these very inclined air showers has evidenced that the emission presents drastically different characteristics compared to vertical showers. In particular, the emission can become incoherent, translating into 
a drop in the electric field strength \cite{Chiche}. In order to understand the specific signatures related to this very inclined regime, we develop phenomenological models and use the Monte-Carlo simulation tools CORSIKA and CoREAS to study the 
electromagnetic content of these air showers, directly linked to the radio signal. We examine the particle populations mainly responsible of the radio emission and the particle distribution in the shower front, taking into account shower-to-shower fluctuations.

\section{Dynamics of particles for very inclined air showers}
\label{section:dynamics}

When ultra high energy particles travel through the atmosphere, they interact with air molecules and produce a cascade of secondary particles, that propagates toward the ground at relativistic speed. The electromagnetic component of the shower is responsible for the production of the radio signal: a beamed coherent emission in the MHz regime. The geomagnetic effect \cite{Scholten} is the major ingredient for radio emission in air: it corresponds to the deflection of electrons and positrons in opposite directions due to the Lorentz force of the Earth magnetic field, balanced by particle scattering with the air molecules. Ultimately, particles are deflected in a plane perpendicular to that of the shower propagation, creating an electric current in the atmosphere.
This current varies with time since the number of electrons and positrons changes during the shower development and the accelerated charges thus induce a  radio signal. The emission is beamed in the forward direction inside a narrow cone due to the relativistic speed of the particles. The half-width of this cone is equal to the Cherenkov angle, typically 1--$2^\circ$ in the air \cite{Schroder}. For an observer located at this specific angle, the radio signals emitted all along the shower arrive simultaneously, boosting the signal along this Cherenkov ring and resulting in brief and high amplitude pulses.

\begin{figure}
\centering 
\includegraphics[width=0.7\columnwidth]{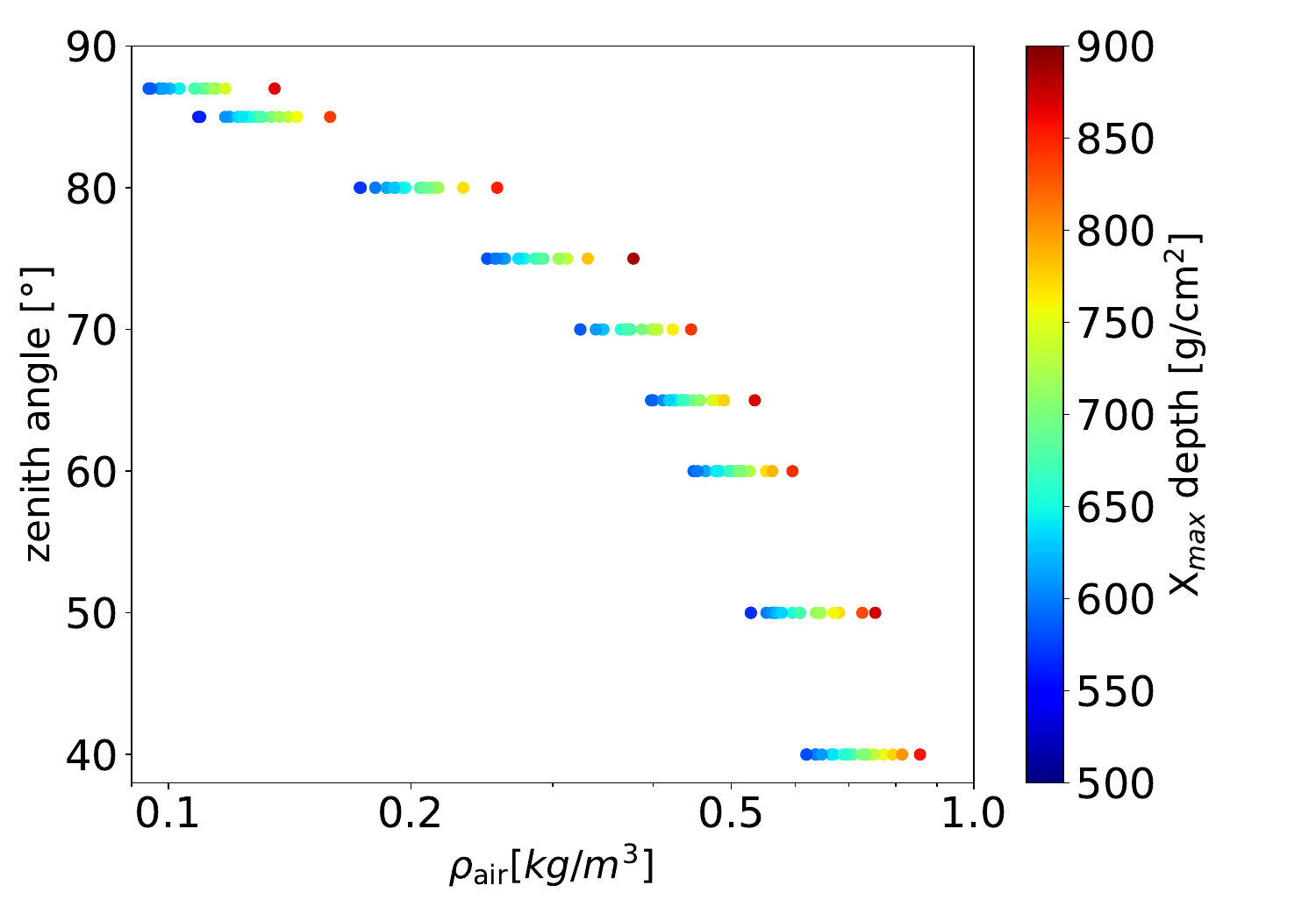}
\caption{Atmospheric density at the maximal shower development for various inclinations. The chosen atmosphere is the U.S. standard atmosphere parameterized according to Linsley~\cite{Hillas}. For a given inclination, the  maximal shower development $X_{\rm max}$ position can be at different depths due to shower-to-shower fluctuations. Very inclined air showers develop in less dense atmospheres.}\label{fig:theta_airdensity}
\end{figure}

Very inclined air showers (with zenith angles $\theta$ larger than $\gtrsim 70^\circ$) present different geometries compared to vertical ones. They can extend over several hundreds of kilometers, as opposed to the $\sim 10$\,km thickness of the troposphere for vertical showers. Moreover, they develop higher in the atmosphere where the air density $\rho_{\rm air}$ is lower (see Figure~\ref{fig:theta_airdensity}). This implies that shower particles have larger mean free paths for collisions with the air molecules, and hence experience more strongly the effect of the Earth magnetic field. 

\subsection{Scattering processes in air showers and deflections by the geomagnetic field}
\label{sec:scattering processes}
Electrons and positrons in the air shower undergo Bremsstrahlung radiation at energies $\epsilon=\gamma m_{\rm e} c^2\gtrsim \epsilon_{\rm c}$, where $\epsilon_{\rm c} = 88\,$MeV is the critical energy below which energy losses due to ionization become dominant --- hence most particles are absorbed in the atmosphere \cite{Engel}.
In the following, we focus on the high energy regime where $\epsilon\gtrsim \epsilon_{\rm c}$ and neglect both ionization and Coulomb scattering, which becomes a dominant process below  $\epsilon_{\rm Coulomb} = 20\,\rm MeV$  \cite{Engel}. The attenuation length for these processes can thus be written \cite{Scholten}: 
\begin{eqnarray}
l_{\rm rad} = \frac{X_0}{\rho_{\rm air}} 
\sim 4\times 10^2\,{\rm m}\,\left(\frac{\rho_{\rm air}}{1\,{\rm kg\,m^{-3}}}\right)^{-1}\ ,
\end{eqnarray}
where $X_0=36.7\,{\rm g}\,{\rm cm}^{-2}$ is the electronic radiation length.  

Electrons and positrons also experience deflections in opposite directions by the geomagnetic field $B$, with Larmor radius 
{$r_{\rm L} = \beta \epsilon {\rm \sin\alpha}/(c e B) \ \sim 6.6\times 10^3 \, {\rm m}\,{\rm \sin\alpha}(\epsilon/100\,{\rm MeV})(B /50\,\mu{\rm T})^{-1}$}, with the reduced velocity $\beta \sim 1$ for relativistic particles, $c$ the speed of light, $e$ the elementary charge and $\alpha$ the so-called geomagnetic angle between the shower arrival direction and the local geomagnetic field direction. In the following, we take ${\rm sin\, \alpha} \sim 1$.

\subsection{Analytical estimate of the air shower lateral extent due to geomagnetic deflection}
\label{sec:analytical_lateralextent}

In this section, we derive an analytical estimate of the particle lateral extent at the shower front, which we define in this section as the lateral separation induced by the geomagnetic field. As explained in Section~\ref{sec:scattering processes}, electrons and positrons in the air shower undergo a circular motion depending on the geomagnetic field $B$ whose radius is the Larmor radius $r_{\rm L}$. The transverse acceleration of electrons yields  $a_t(t) = {{\rm d} v_t(t) }/{{\rm d}t} = {\beta^{2}c^{2}}/{r_{\rm L(t)}} $ which in the limit $\beta \sim 1$ gives ${{\rm d} v_t(t) }/{{\rm d}t} = {c^{3}e B}/{\epsilon}$. Assuming a continuous Bremsstrahlung particle energy loss in the air of electrons with initial energy $\epsilon_0$, we get $\epsilon(t) = \epsilon_0 e^{-t/\tau}$. The transverse particle acceleration in the shower front at time $t$ thus reads: 
\begin{equation}
\frac{{\rm d}^2x_{\rm lat}}{{\rm d}t^{2}}=\frac{c^3eB}{\epsilon_0\, \exp(-t/\tau)}\,
\end{equation}
where $x_{\rm lat}$ is the particle transverse position (orthogonal to the shower axis) and $\tau = l_{\rm rad}/c$ the Bremsstrahlung energy loss timescale \cite{Scholten}. It yields a transverse velocity: 
\begin{eqnarray}
\nonumber
    v_t(t) = \frac{{\rm d}x_{\rm lat}}{{\rm d}t} = \int_0^t \frac{{\rm d} v_t(t') }{{\rm d}t'} {\rm d}t' &=&  \int_0^t  \frac{c^{3}e B}{\epsilon_0 e^{-t'/\tau}} {\rm d}t' \  \\
   &=& \frac{c^{3}\tau e B}{\epsilon(t)}\left( 1- \frac{\epsilon(t)}{\epsilon_0}\right).
\end{eqnarray}
where $t=0$ corresponds to the creation of the particle.
Integrating over time, we finally obtain the transverse position of particles in the shower as a function of time \cite{Chiche}:  
$x_{\rm lat}(t) =  \tau^{2}c^{3}eB\rm \, (e^{t/\tau} -1 -{t}/{\tau})/ \epsilon_{0}$.
The shower lateral extent then reads:
\begin{eqnarray}
\label{eq:lateral_extent}
\begin{split}
l_{\rm lat} = 2x_{\rm lat}(t = \tau) =&  2\left(\frac{l_{\rm rad}}{c}\right)^{2} \frac{c^{3}eB (\rm e^{1} - 2)}{\epsilon_{0}} \  \\
\sim&  3\times10^{1}\,{\rm m}  \left(\frac{\rho_{\rm air}}{1\,{\rm{ kg\,m^{-3}}}}\right)^{-2} \left(\frac{B}{50\,\rm \mu \rm T}\right) \left(\frac {\epsilon_{0}}{100\,\rm MeV}\right)^{-1} \ .
\end{split}
\end{eqnarray}
where the factor 2 accounts for the dynamics of both positrons and electrons.


This lateral extent is thus specifically the extent induced by the geomagnetic field for particles at high energy regime, which are the main particles involved in the radio signal as shown in Section~\ref{sec:contribution}.

As demonstrated in Section~\ref{sec:radio_point}, most of the emission is produced in the region around the shower axis where the electron/positron density is the higher, and more specifically at $X_{\rm max}$. The amplitude of the radio signal scales linearly with the number of electrons and positrons and its power scales quadratically.

If the shower lateral extent at $X_{\rm max}$ is larger than the spatial coherence length, the signal becomes incoherent \cite{Chiche}. 

\subsection{Analytical estimate of the air shower lateral extent over the full zenith angle range}
\label{sec:full_analytical_lateralextent}

When the radiation losses dominate over the geomagnetic deflection, the Molière radius describes the lateral spread of the particles in the air shower. This extension, due to Coulomb scattering principally, can be expressed as $R_{\rm M} = (9.6\,\rm g/cm^{2})/\rho_{\rm air}$~\cite{Scholten}. 

Two distinct regimes can then be highlighted for low and high air shower inclinations, corresponding to high and low air densities respectively: while the lateral extent of vertical air showers is driven by multiple scattering, scaling in $\rho_{\rm air}^{-1}$, very inclined air showers experience a drastic lateral extent increase due to the Earth's magnetic field and scale in $\rho_{\rm air}^{-2}$. 

The shower lateral extent over the full zenith angle regimes and for all frequencies can then be expressed analytically as a broken power-law:
\begin{eqnarray}
\label{eq:lateral_extent_full}
l_{\rm lat} \sim \left\{
    \begin{array}{ll}
     30\,{\rm m}  \left({\rho_{\rm air}}/{1\,{\rm{ kg\,m^{-3}}}}\right)^{-2} \left({B}/{50\,\rm \mu \rm T}\right) \left({\epsilon_{0}}/{100\,\rm MeV}\right)^{-1}  &\quad  \mbox{for } \rho_{\rm air}\lesssim \rho_{\rm air,c}\\
     96\,{\rm m} \, ({\rho_{\rm air}}/{1\,{\rm{ kg\,m^{-3}}}})^{-1} &\quad \mbox{for } \rho_{\rm air}\gtrsim \rho_{\rm air,c}
    \end{array}
\right. \ ,
\end{eqnarray}
where $\rho_{\rm air, c}\sim 0.3\,({\rho_{\rm air}}/{1\,{\rm{ kg\,m^{-3}}}}) \left({B}/{50\,\rm \mu \rm T}\right)^{-1} \left({\epsilon_{0}}/{100\,\rm MeV}\right)$ {$\rm kg\,m^{-3}$} .

In the following, we examine numerically, with Monte-Carlo microscopic simulations, the content of particle air showers, and demonstrate that this theoretical expression is indeed verified.

\section{Contribution of air shower particles in radio emission}
\label{sec:contribution}
One of the first steps to link the particle content of very inclined EAS with the specific features of the radio signal consists in identifying the particles that are specifically involved in the radio emission. For this purpose, we will rely on microscopic numerical simulations. 

\subsection{Microscopic simulations with CORSIKA and CoREAS}\label{section:sims}

\begin{figure}
\centering 
\includegraphics[width=0.7\columnwidth]{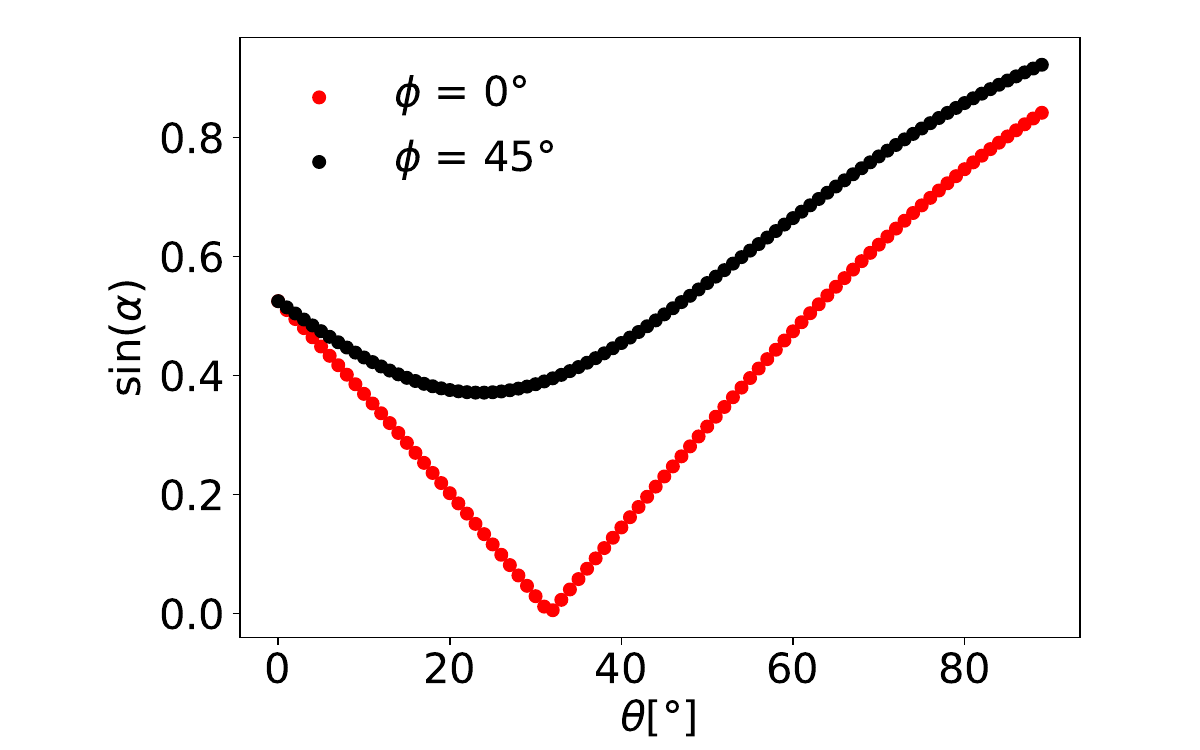}
\caption{Values of $\rm sin\,\alpha$ covered by our set of CORSIKA simulations, as a function of the zenith angle $\theta$. The geomagnetic angle $\alpha$ corresponds to the angle between the shower axis and the magnetic field, set to the value of the Dunhuang magnetic field $B_{\rm Dunhuang}=56\,\mu T$.}
\label{fig:geomagnetic_angle}
\end{figure}

We simulate air showers with CORSIKA 7.7500 \cite{CORSIKA}, a detailed Monte-Carlo code which follows the individual trajectories of the particles. Quantum variations on $X_{\rm max}$, between air showers with identical primaries, are intrinsically modelled. 
QGSJetII is chosen as the high-energy interaction model and URQMD 1.3.1 as the low-energy interaction model. The primary particle is a proton with an energy of $10^{17}$\,eV, with a zenith angle which can vary between $0^\circ$ and $87^\circ$ throughout the study depending on the simulation sets chosen and two possible azimuth angles ($0^\circ$, $45^\circ$). In CORSIKA conventions, these correspond to down-going air showers, $\theta=0^\circ$ being a perfectly vertical air shower and $\phi=0^\circ$ 
is a shower coming from South and going to North, as the azimuth angle is defined in the shower propagation direction. The elevation site (1140 m) and magnetic field correspond both to those of a site similar to Dunhuang, China, where the GRAND prototype is being deployed \cite{Joao}. The atmosphere is approximated with the standard Linsley model \cite{Hillas}. Figure~\ref{fig:geomagnetic_angle} shows the range of values covered by $\rm sin\,\alpha$ with these specific sets of parameters. The effective magnetic field $B_{\rm eff} = B\,{\rm sin}\,\alpha$ is close to $B$ for very inclined air showers in these configurations.

We store the information on the particles of the EAS (position, energy) that cross the shower plane at $X_{\rm max}$ 
({if not otherwise specified}) thanks to the inclined plane option in CORSIKA. The shower plane is the plane perpendicular to the direction of propagation of the air shower. In this plane, the axes are usually aligned with the Lorentz force and thus with the $({\bf v}\times {\bf B})$ direction and with the $[{\bf v}\times ({\bf v}\times{\bf B})]$ to get an orthogonal coordinate system. 

As the parameter $X_{\rm max}$ depends on intrinsic and detailed behavior of the particles while they propagate through the atmosphere, it can be accessed only once the simulation is completed. However, the value of $X_{\rm max}$ is required prior to running the simulation, in order to set the shower plane position where particle information is recorded. We therefore run each simulation twice with the same seed to avoid fluctuations: first to find the value of $X_{\rm max}$ and a second time to store the particle parameters at this position.

For the radio emission computation, we use CoREAS V1.4 \cite{CoREAS}, the extension of CORSIKA that allows to simulate the electromagnetic radiation produced by the particles of the shower. The same input parameters as CORSIKA are used, and antennas are located on the ground, on the North axis at a fixed position according to the zenith angle. This position is chosen to be around –-and not strictly at-- the maximum of the radio signal on this axis. 
The time traces obtained thanks to CoREAS are used to define the contribution of different shower stages and particle energy regimes in the emission of the radio signal.  


\begin{figure}[htb]
\centering
         \includegraphics[height=0.5\columnwidth]{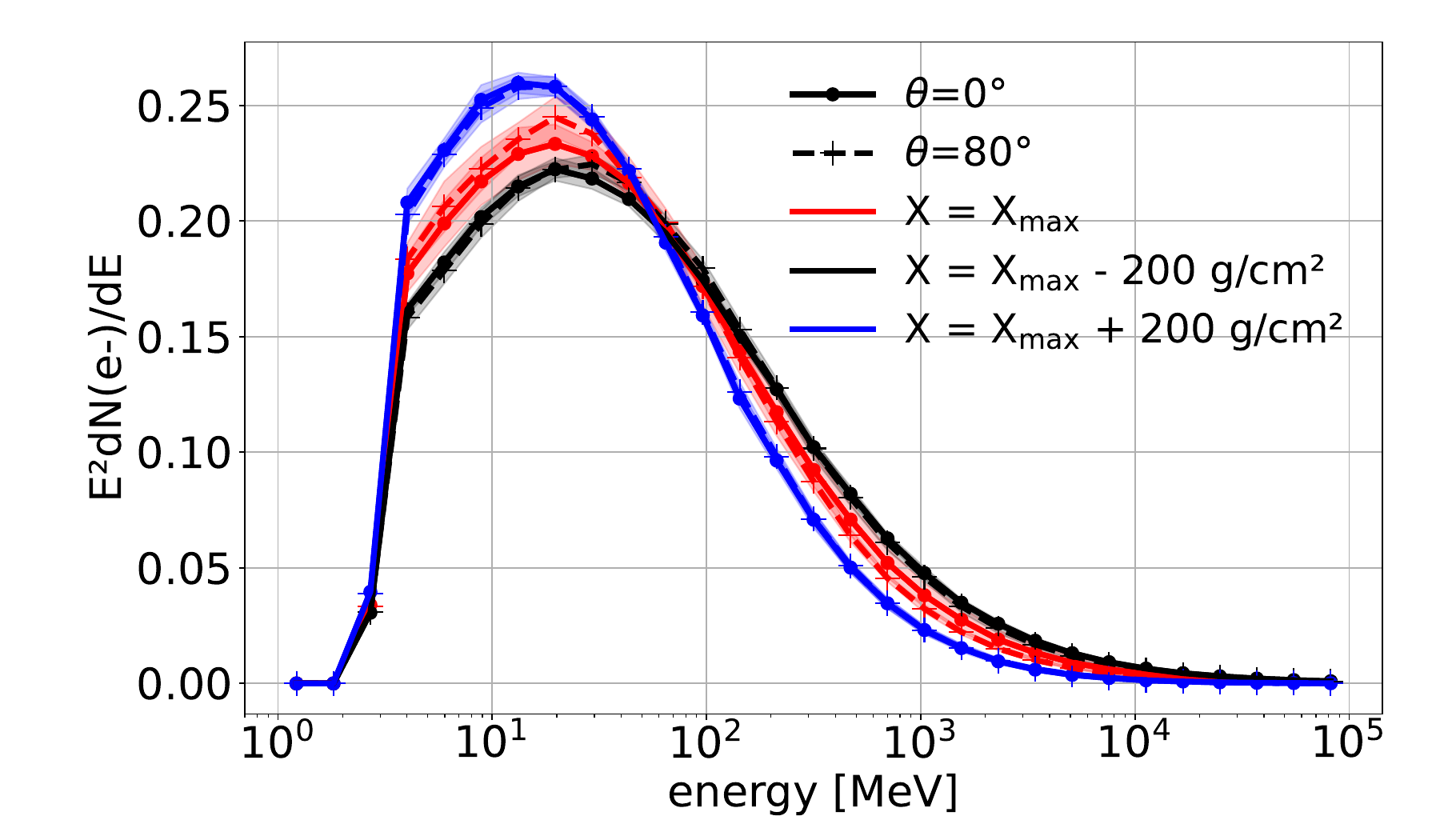}
     \caption{{
     Electrons normalized energy spectra at different shower depths ($X$) and inclinations ($\theta).$} The primary particle is a proton with energy $10^{17}\,\rm eV$ with $\theta = 0^\circ$ (dashed lines) and  $\theta = 80^\circ$ (lines). The shaded regions indicate the 
     width of the $\sigma$ distribution for each energy bin due to statistical fluctuations over 40 showers for each zenith angle.}
      \label{fig:energy_spectrum}
\end{figure}

\subsection{Energy spectra of the 
electromagnetic component of air showers}
\label{sec:spectra}
We study the particle energy content in air showers at fixed primary energies and different 
inclinations with CORSIKA simulations. We store the energy of electrons and positrons of the air showers when they cross the shower plane 
at different stages of the shower development. 

The energy spectra of electrons for two different zenith angles are presented in Figure~\ref{fig:energy_spectrum}. 
The colored lines and dashed lines indicate the total energy of particles contained in each energy bin, at a given atmospheric depth $X$ of the shower for zenith angles $0^\circ$ and $80^\circ$ respectively. The depth $X$ refers to a physical location along the longitudinal axis of the air shower depending on the amount of crossed atmosphere. 
At early stages (with lower shower development stage $X$ and so at less amount of crossed atmosphere), the shower contains higher energy particles than at later stages. This feature is expected as these particles lose energy during the shower development. 
Furthermore, the particle content is similar for vertical and very inclined air showers at a given depth, with a peak energy in the range [10, 100] MeV at the maximal shower development $X_{\rm max}$, which is around the critical energy $\epsilon_{\rm c}$.


\subsection{Contribution of particle of different energies to radio emission}
\label{sec:energy}

\begin{figure}[!tb]
     \centering
         \includegraphics[height=0.5\columnwidth]{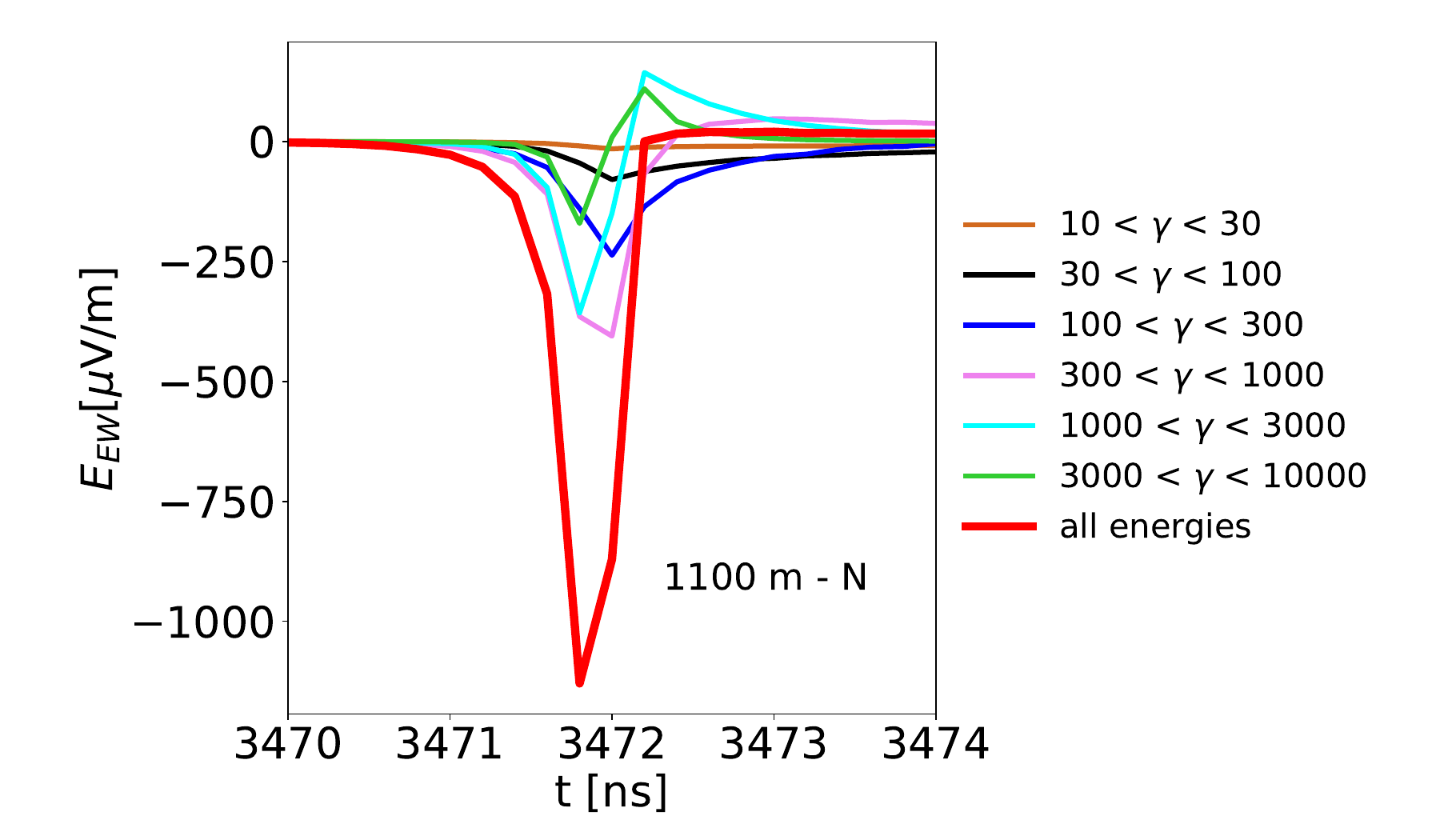}
      \caption{Traces of the radio signal created at different electron/positron energy (Lorentz factor) ranges for an air shower with inclination $\theta=70^\circ$ and azimuth $\phi=0^\circ$. The primary particle is a proton with energy $10^{17}\,\rm eV$. The antenna is located at a distance of 1100\,m of the shower core on the North axis, where the amplitude of the radio signal is maximum.The lines represent the electric fields produced by particles in specific energy ranges as indicated.}
    \label{fig:trace70}
\end{figure}

\begin{figure}[h]  
     \centering
     \includegraphics[height=0.5\columnwidth]{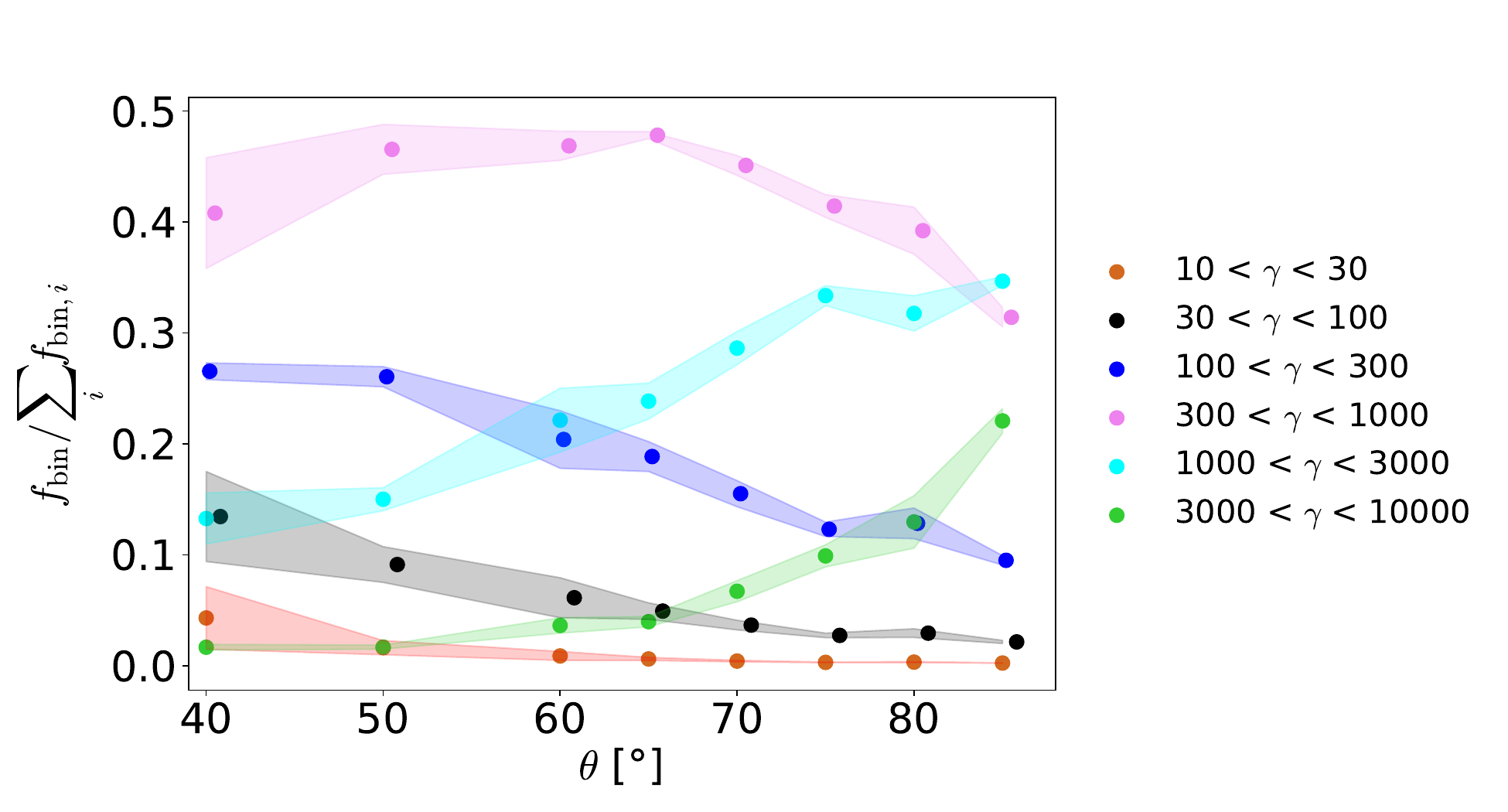}
     \includegraphics[height=0.5\columnwidth]{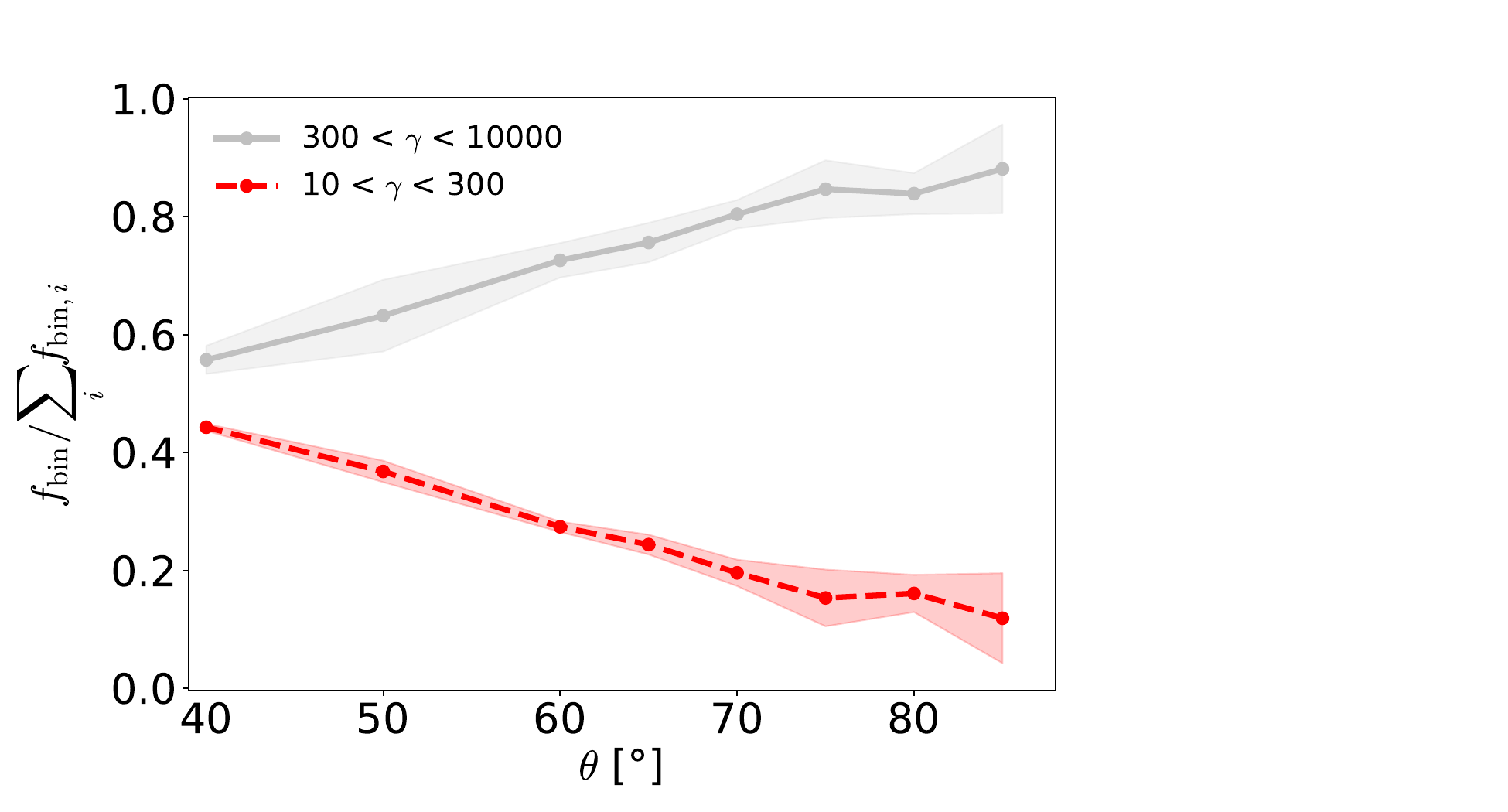}
     \caption{(Top) Contribution of different particle energy regimes responsible for the radio signal. Each point corresponds to the mean value of the normalized energy fluence in a given particle energy range on a set of 40 simulations. The shaded regions represent the 
     width of the distribution $\sigma$. 
    (Bottom) Sum of the contribution of particle energy regimes for $ 10 < \gamma <300 $ and $ 300 < \gamma < 10000$, with corresponding shaded 
    regions.}
      \label{fig:energy}
\end{figure}

In the next step, we examine the contribution of particles of various energies to the radio signal. In this Section, we use raw data without any frequency filtering, in order to study this contribution across the entire signal. In Appendix~\ref{sec:frequency}, we present this study on data filtered in the frequency range [50, 200] MHz. Radio antennas are indeed sensitive to a limited bandwidth of the electromagnetic spectrum and several experiments are constrained in this frequency range. We find that frequency filtering does not affect our results. 

We run CoREAS simulations for a fixed primary with different zenith angles, and record the electromagnetic radiation emitted by the particles of the air showers in specific energy bins ---i.e., in specific Lorentz factor bins--- and for the whole range of energy (similar methods are presented in \cite{Huege_2007, 2011PhDT156L}). In Figure \ref{fig:trace70}, we represent the traces of the radio signal for a zenith angle of $70^\circ$. 
Only the East-West polarisation of the electric field is represented, as it is largely dominant in this specific configuration: for an air shower with azimuth angle $\phi = 0\,^{\circ}$ and an antenna located on the North axis, the North-South polarization of the electric field is negligible compared to the East-West polarization. 

The radiation energy, which is defined as the energy emitted by the air shower in the form of radio waves, comes from the radiation generated by the electromagnetic part of the air shower \cite{Glaser}. To determine this radiation energy, one method is to interpolate and integrate the measured energy fluence on the ground \cite{energy_radiation}. This energy fluence $f$, which is the energy deposited per area, can be expressed as: 

\begin{eqnarray}
\label{eq:energy_fluence}
\label{eq:energy_fluence_2}
f = \epsilon_{0}c\Delta t \sum_{i} (E^{2}_{\rm EW}({\bf  r}, t_{i} )+E^{2}_{\rm NS}({\bf  r}, t_{i} )+E^{2}_{\rm vertical}({\bf  r}, t_{i}))\ ,
\end{eqnarray}
with $\epsilon_{0}$ the vacuum permittivity, $c$ the speed of light 
and $\Delta t$ the sampling interval of the electric field ${\bf E}({\bf r}, t)$ which is broken down into three components: the East-West polarization $E_{\rm EW}({\bf r}, t)$, the North-South polarization $E_{\rm NS}({\bf r}, t)$ and the vertical polarization $E_{\rm vertical}({\bf r}, t)$

We calculate the energy fluence created by each energy bin $f_{\rm bin}$ and received by the antenna as presented in~Eq~\ref{eq:energy_fluence_2}. We normalize this result by dividing it by the sum of the energy fluence of each pulse $\sum_{j} f_{\rm bin,j}$ and we obtain: $f_{\rm bin}/\sum_{j} f_{\rm bin,j} = \sum_{i} E_{\rm bin}^{2}(t_{i})/ \sum_{j} (\sum_{i} E_{\rm bin,j}^{2}(t_{i}))$ with $E_{\rm bin}^{2} = E_{\rm EW, bin}^{2} + E_{\rm NS, bin}^{2} + E_{\rm vertical, bin}^{2}$.  
This enables us to compare the contribution of the different particle populations to the radio signal for each zenith angle.

Figure~\ref{fig:energy} (top) shows the mean values and uncertainties of this fraction obtained for a set of 40 simulations on each zenith angle to take into account the shower-to-shower fluctuations.
It appears that the electric field is dominated by contributions from increasingly energetic particles as the inclination increases. It is highlighted by Figure~\ref{fig:energy} (bottom) that represents the sum on the mean values split into two energy ranges: $10 < \gamma < 300$ and $ 300 <\gamma<10000$. For inclinations larger than $70^\circ$, the contribution from low energy particles is less important, representing less than 20\% of the total contribution.

We can therefore propose two energy regimes at low and high (> 70°) inclinations, which are approximate guesses motivated by the general trends observed in Figure 5 (top).
We 
therefore consider that the electric field is roughly dominated by [50, 1500] MeV particles 
for very inclined air showers, and energy losses are mainly due to Bremmstrahlung in this energy range.
For showers with a lower inclination, another energy regime is dominant, in the range [10, 500] MeV.

Using this method, we estimate the contribution of the radio emission at ground from particles with different energies. But because we measure the energy fluence of each energy bin independently, we ignore possible destructive interferences between the emission of particles from different energy bins, and we overestimate the total electric field. 
Another method is proposed in Appendix~\ref{sec:interferences} to examine this effect. The trends obtained are identical, hence both treatments lead to the conclusion presented above.

\subsection{Radio emission region in the air shower}
\label{sec:radio_point}

\begin{figure}[!tb]
     \centering
         \includegraphics[height=0.5\columnwidth]{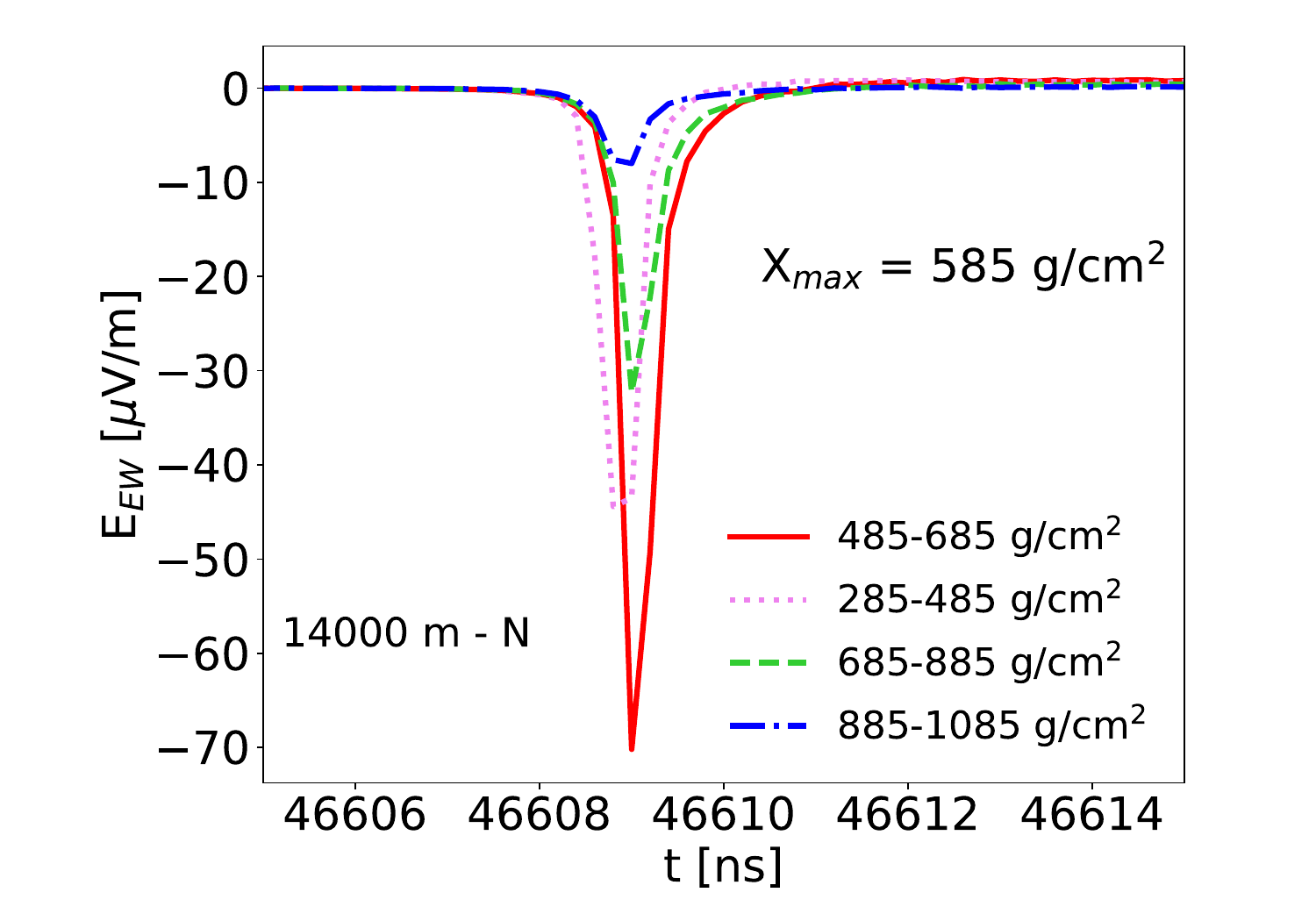}
         \caption{Traces of the radio signal created at different stages of the shower development by particles in the energy range 100 < $\gamma$ < 3000  ([50, 1500] MeV). The primary particle is a proton with energy $10^{17}\,\rm eV$, zenith angle $\theta = 85^\circ$ and azimuth $\phi = 0^\circ$. The antenna is located on the North axis at a distance of 14000\,m from the shower core, where the radio signal is maximum.}
         \label{fig:slantdepth_theta85}
     \end{figure}

\begin{figure}[!htb]
    \centering
        \includegraphics[height=0.5\columnwidth]{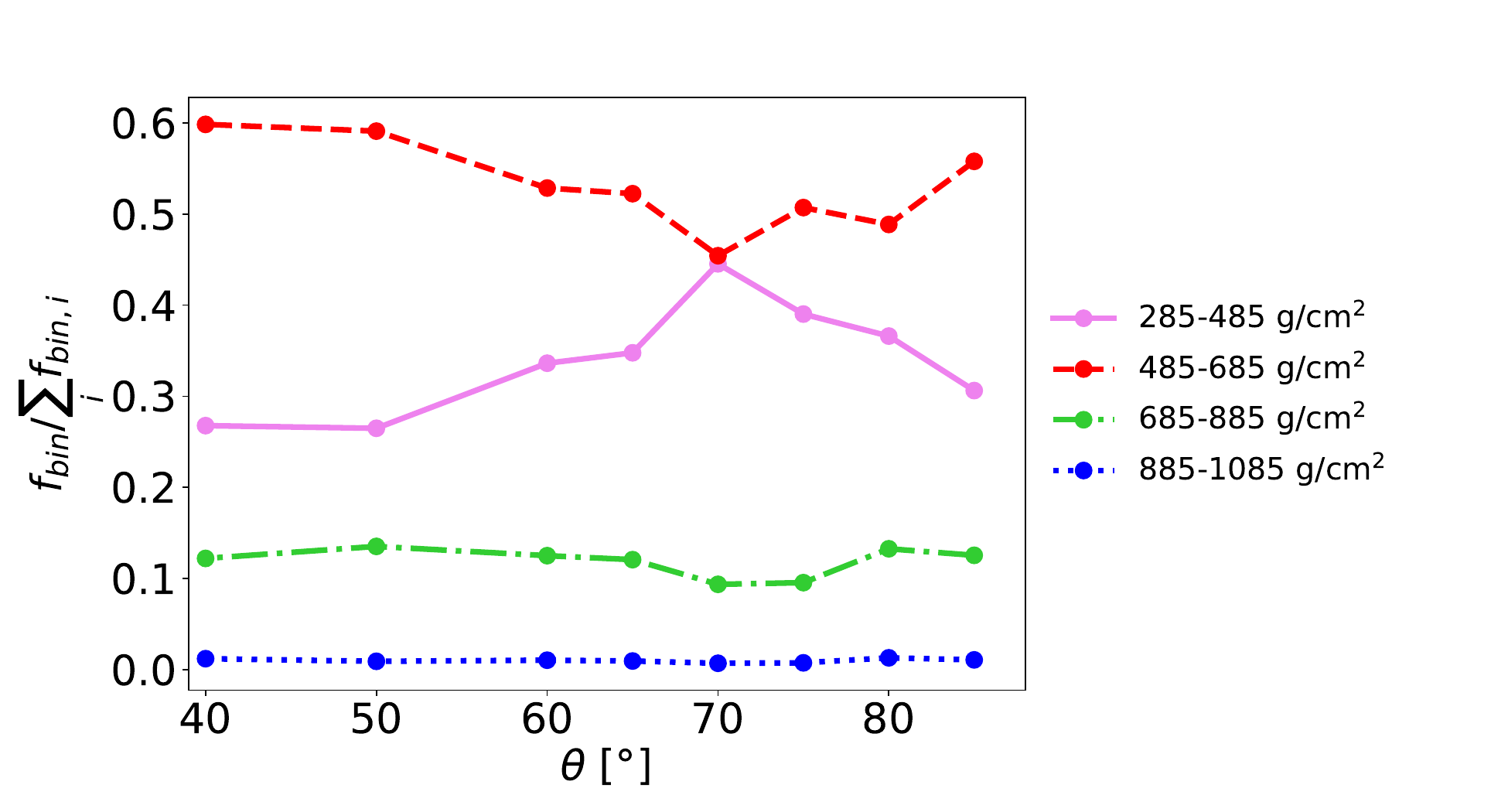}
         \label{fig:slantdepth_theta80}
     \caption{Contribution of shower stages in radio emission. Each point corresponds to the normalized energy fluence 
     in a given slant depth range. For all the simulations,the seed is fixed so that the value of the maximum shower development is the same for all inclinations: $X_{\rm max} = 585\,\rm g/cm^{2}$.}
      \label{fig:slantdepth}
\end{figure}

The contribution of different shower phases can also be 
estimated for air showers with different zenith angles. Following the same procedure as before, we generate with CoREAS the electric field emitted by particles with energy in the range [50, 1500] MeV or [10, 500] MeV according to the zenith angle (see Section~\ref{sec:energy}). In this specific study, the air shower is divided in the simulation into segments of atmospheric depths and each radio pulse corresponds to the signal emitted by the corresponding segments. Figure~\ref{fig:slantdepth_theta85} represents the time traces of the East-West polarisation of the electric field for $\theta = 85^\circ$ and an antenna located on the North axis where the radio signal is maximum.
Comparing the normalized energy fluences in each bin of atmospheric depth for different zenith angles 
as shown in Figure~\ref{fig:slantdepth}, we can see that the most representative region of the radio emission is around the maximal development of the shower $X_{\rm max}$ whatever the inclination. This confirms the commonly admitted notion that the radio emission can be assumed to be produced at $X_{\rm max}$, also for very inclined air showers~\cite{Deconene2020}.

\section{Particle distribution and lateral extension of very inclined air showers}
\label{lateral_extent}

In the previous sections, we found that the energy spectrum in the shower plane is similar for vertical and very inclined air showers but that most energetic particles contribute to the total radio emission for very inclined showers.
The identification of particles responsible for the radio signal allow us to study the particle content of air showers in the relevant energy ranges and to highlight the distinct features on the particle distribution for different inclinations. We also found that the shower development stage $X$ that produces the radio signal does not vary with the zenith angle. The robustness of this last parameter can be used to give a universal definition of the lateral extent of electrons and positrons in the air shower in two energy regimes depending on the inclination.

\begin{figure}[!tb]
     \centering
         \includegraphics[height=0.5\columnwidth]{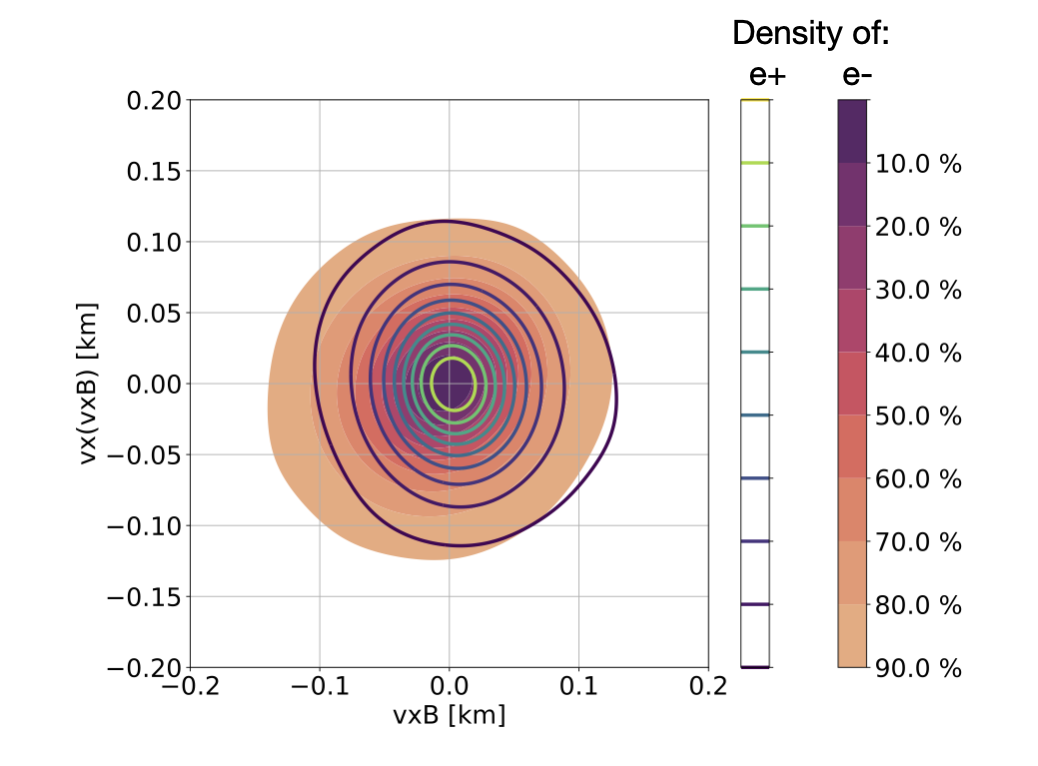}
         \includegraphics[height=0.5\columnwidth]{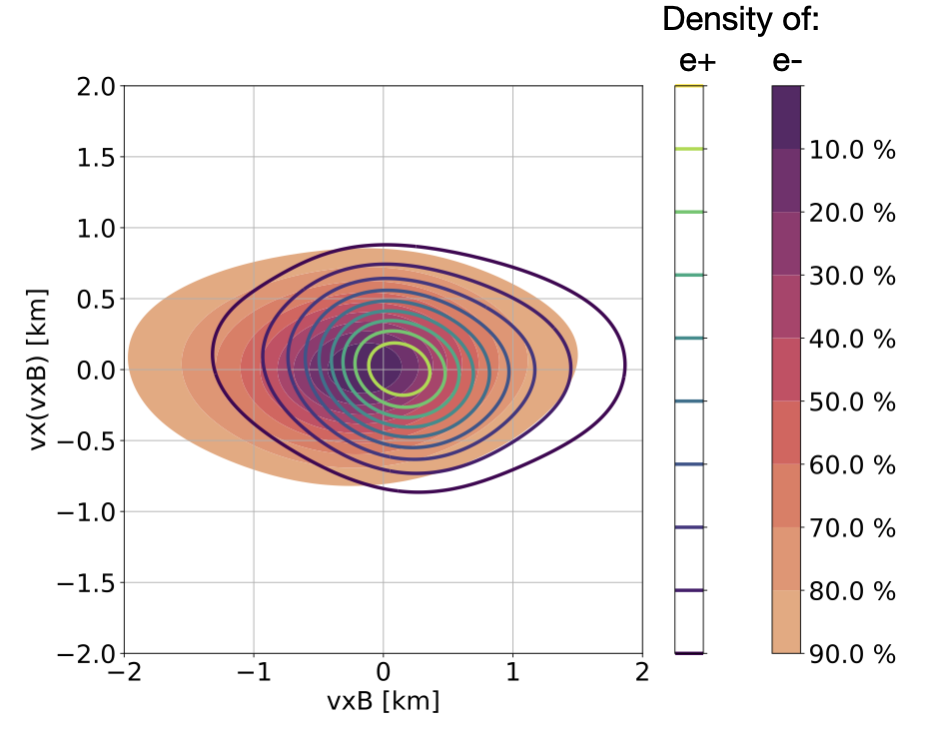}
     \caption{Density of electrons and positrons in the shower plane at $X_{\rm max}$. The color (linear) contours delimit the region within which the number of electrons (positrons) divided by their total number in the shower at $X_{\rm max}$ within the selected energy range as indicated. The primary particle is a proton with energy $10^{17}$ eV and azimuth $\phi = 0^\circ$. (\textit{Top}) $\theta = 60^\circ$ and only particles in the energy range [10, 500] MeV are selected. (\textit{Bottom}) $\theta = 87^\circ$ and only particles in the range [50, 1500] MeV are selected. 
     }
      \label{fig:densityplot_twoangles}
\end{figure}

\subsection{Particle distribution of air showers for various inclination regimes}
\label{sec:distribution}

\begin{figure}[!tp]
     \centering
         \includegraphics[height=0.4\columnwidth]{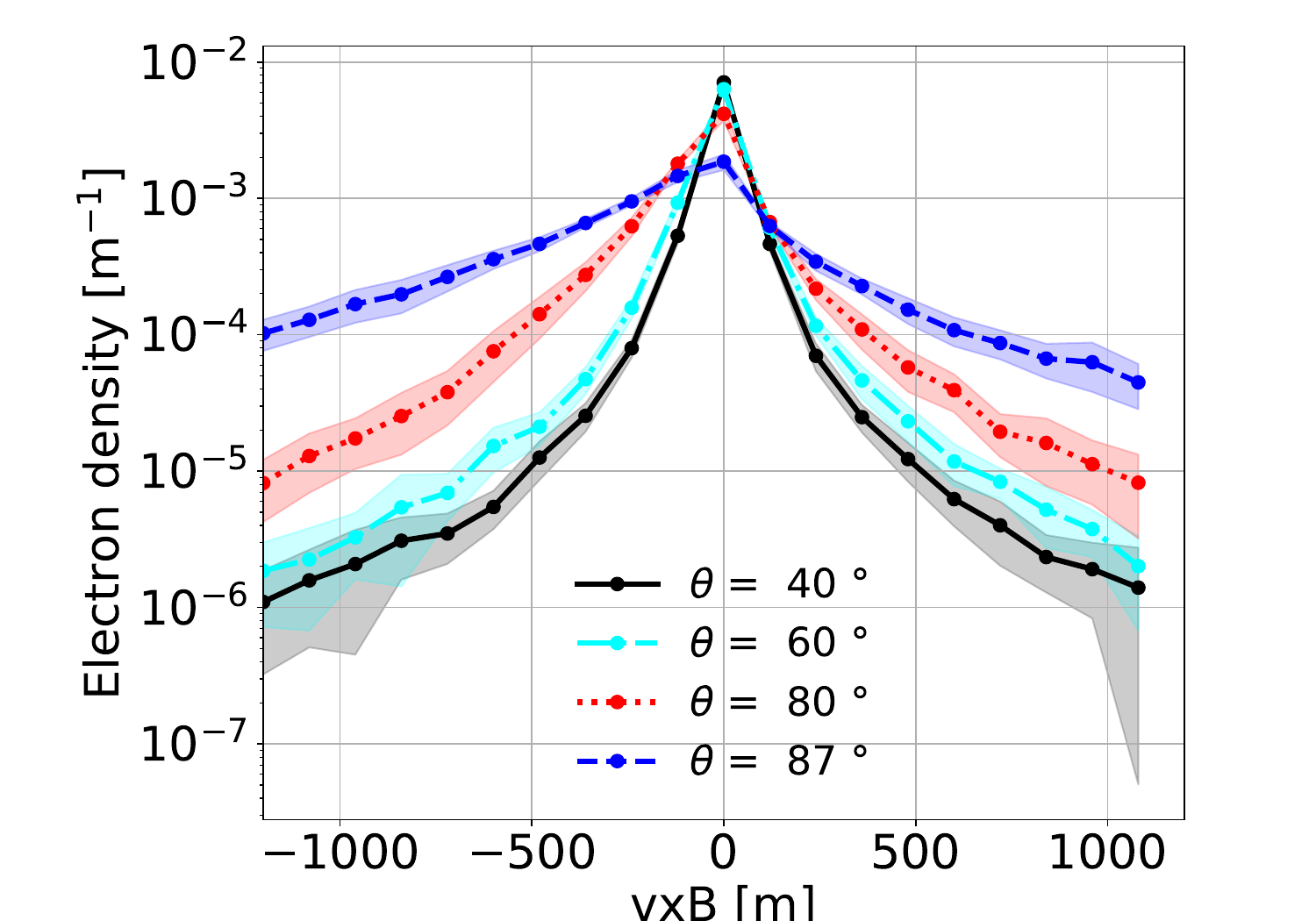}
         \includegraphics[height=0.4\columnwidth]{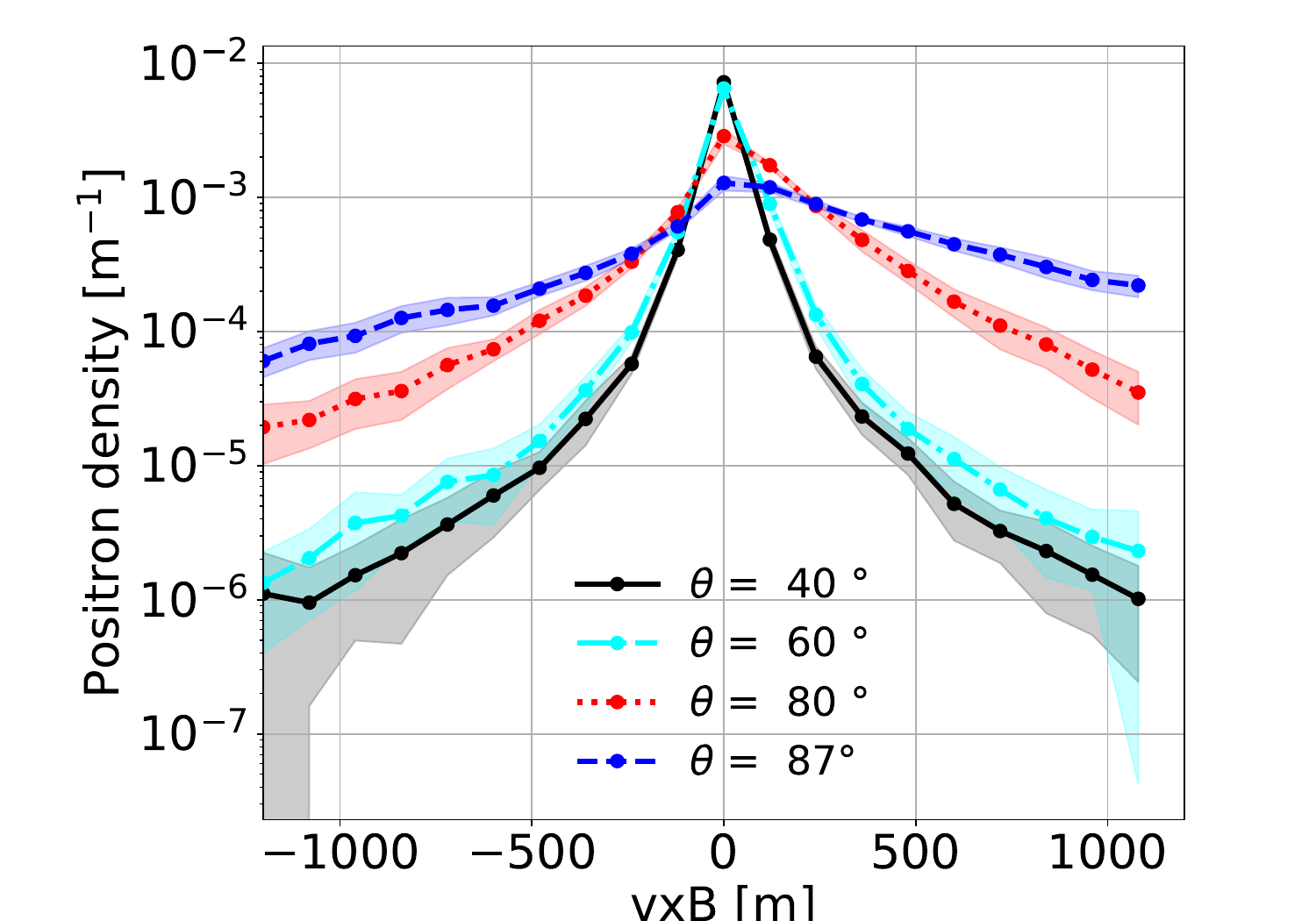}      
         \includegraphics[height=0.4\columnwidth]{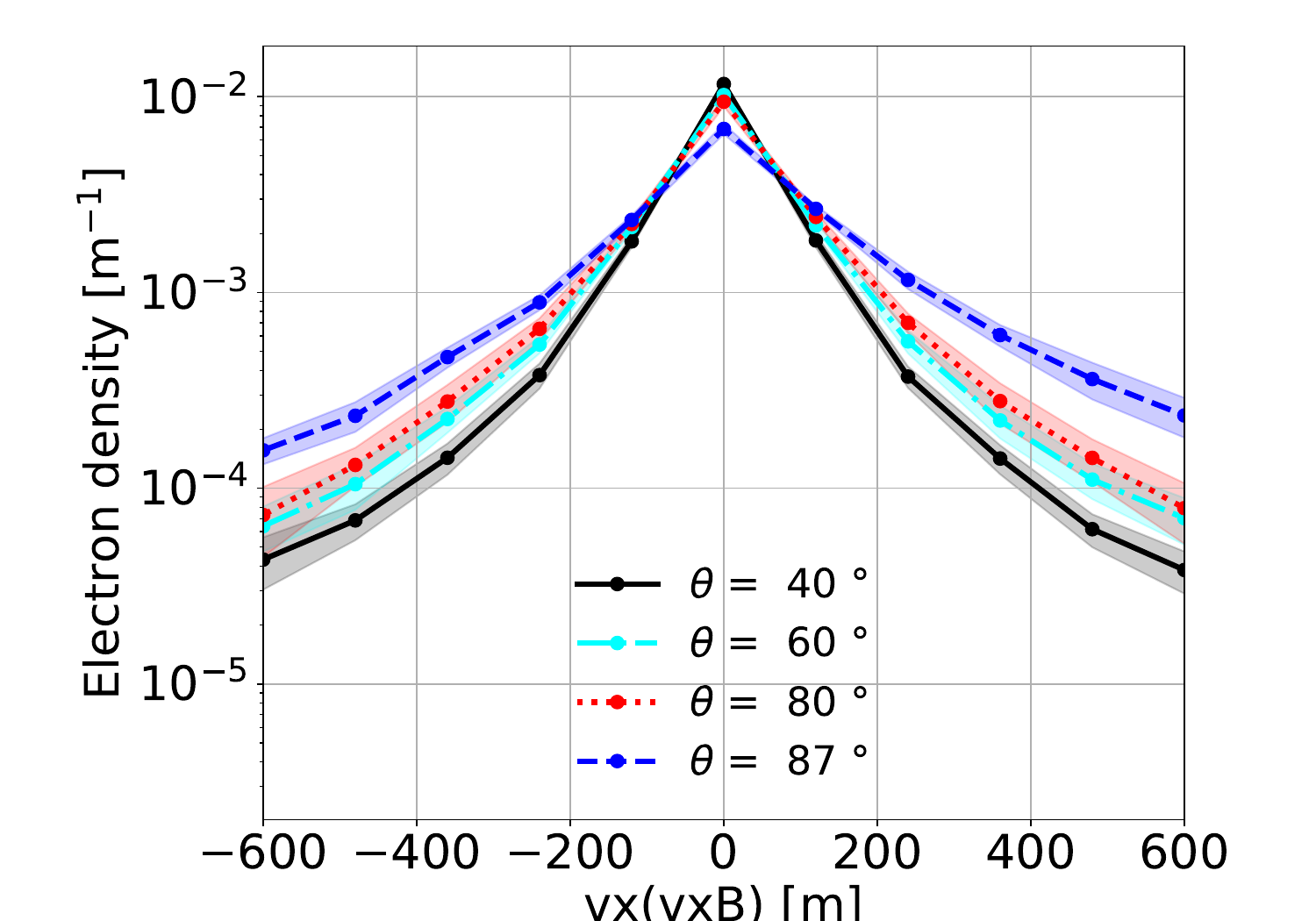}   
     \caption{ Electron (top) and positron (middle) distributions projected on the $({\bf v}\times {\bf B})$ and $({\bf v}\times [{\bf v}\times{\bf B}])$ (bottom) axes at $X_{\rm max}$ for various inclinations. The primary particle is a proton with energy $10^{17}$\,eV. For $\theta = 80^\circ$ and $\theta = 87^\circ$ ($\theta = 40^\circ$ and $\theta = 60^\circ$), only particles in the energy range [50, 1500] MeV ([10, 500] MeV) are selected. Each point corresponds to the mean value of the number of electrons or positrons on a set of 50 simulations for each zenith angle and the shaded regions correspond to the corresponding 
     width of the distribution.}
      \label{fig:distribution_along_distance}
\end{figure}

We study the lateral distribution of electrons and positrons in the shower plane $[{\bf v}\times {\bf B}, {\bf v}\times ({\bf v}\times{\bf B})]$ at $X_{\rm max}$ with CORSIKA simulations, $\bf v$ being the shower axis and $\bf B$ the local magnetic field direction. Figure~\ref{fig:densityplot_twoangles} represents the 2D particle density contained inside 
isocontours in the shower plane for a vertical ($\theta = 60^\circ$) and a very inclined ($\theta = 87^\circ$) air showers in specific energy ranges (see Section~\ref{sec:energy}). For both zenith angles, most of the particles are distributed at the center of the shower, where most secondary particles are produced, but the spatial distribution look different. For the vertical air shower, electrons and positrons are isotropically dispersed around their barycenter. The charge separation along the $({\bf v}\times {\bf B})$ axis, 
which is a deflection in opposite directions as mentioned in Section~\ref{section:dynamics} under the influence of the Lorentz Force $\bf F = \pm q{\bf v}\times {\bf B}$, is concentrated in the first meters around the shower axis of coordinates $(0,0)$. In this inclination regime, the particle positions are constrained within the Molière radius $R_{\rm M}$ given by the multiple Coulomb scatterings from high air density \cite{PhysRevD.98.030001}: there is therefore a competition between Coulomb diffusion and charge separation. At a height of 4 km (typical height of the shower maximum for vertical air showers), the atmospheric density is $\rho_{\rm air} \sim 0.82\,{\rm kg\,m^{-3}}$ that yields to $R_{\rm M} \sim 117\,{\rm m}$ \cite{Scholten}. For the very inclined air shower, the charge separation between electrons and positrons is stronger and coupled with a drastic lateral extent increase in the $({\bf v}\times {\bf B})$ direction. The particle positions are not constrained inside the Molière radius, which has a typical value of $R_{\rm M} \sim 800\, {\rm m}$ for a shower with a zenith angle of 87° and where the air density at the maximal shower development is around $\rho_{\rm air} \sim 0.12\, {\rm kg.m^{-3}}$.  It is a regime where air showers propagate in less dense atmosphere, resulting in a higher mean free path between collisions and particle deflection dominates, giving anisotropic particle distributions.

These specific features are highlighted in Figure~\ref{fig:distribution_along_distance} which represents the 
density of electrons and positrons projected along the $({\bf v}\times {\bf B})$ or $[{\bf v}\times ({\bf v}\times{\bf B})]$ axis for $\theta = 40^\circ, 60^\circ, 80^\circ, 87^\circ$ on a set of $100$ simulations. While the lateral extent clearly drastically increases in the $({\bf v}\times {\bf B})$ direction for very inclined air showers (there are still 10\% of electrons at $1000$\,m from the shower core in the $-({\bf v}\times {\bf B})$ direction for a very inclined air shower for only 0.02\% of electrons at the same distance for a vertical air shower), their charge separation is also larger. 
Along the $[{\bf v}\times ({\bf v}\times{\bf B})]$ direction, the size of the shower front does not depend on deflection induced by the geomagnetic field: particles are thus isotropically dispersed around the shower axis for all inclinations. 

\subsection{Energy distribution in the shower plane}
\begin{figure}[!h]
         \includegraphics[width=0.5\columnwidth]{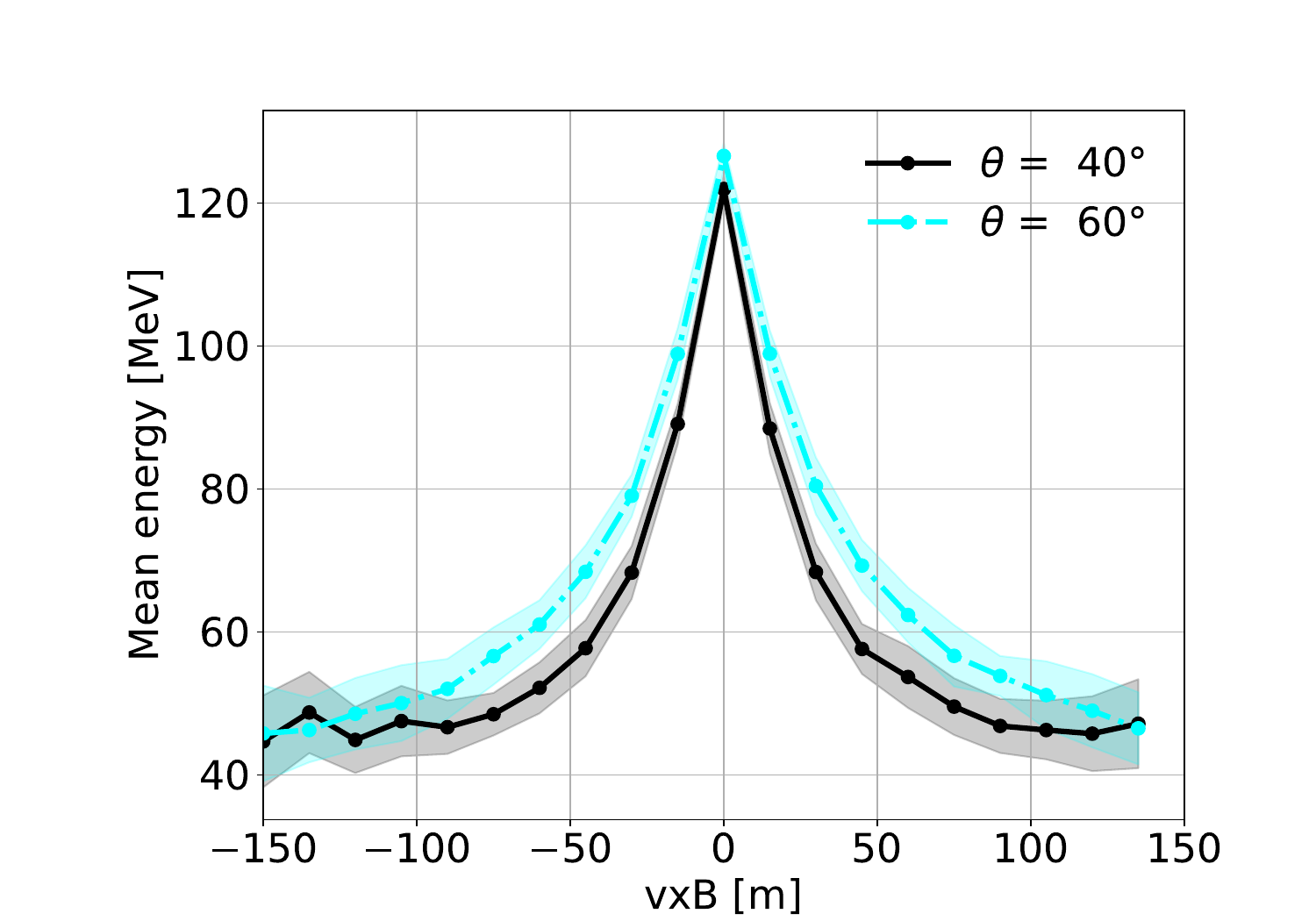}
         \includegraphics[width=0.5\columnwidth]{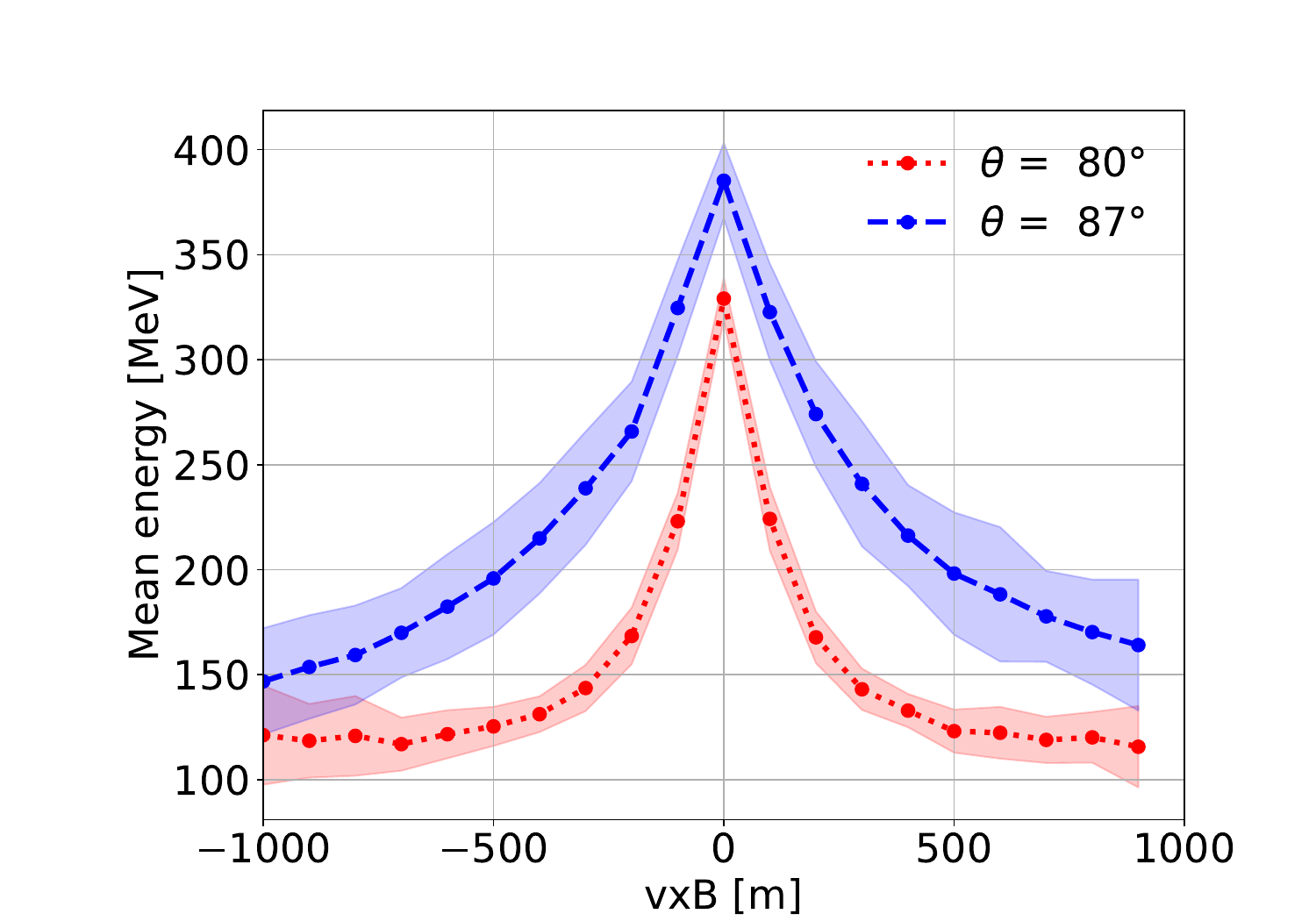}
         
     \caption{Electrons/positrons mean energy distribution on the $({\bf v}\times {\bf B})$ axis at $X_{\rm max}$ for various inclinations. (\textit{Left}) For vertical showers with $\theta = 40^\circ$ and $\theta = 60^\circ$ where only particles in the energy range [10, 500] MeV are selected. (\textit{Right}) for very inclined air showers with $\theta = 80^\circ$ and $\theta = 87^\circ$ where only particles in the energy range [50, 1500] MeV are selected. The shaded regions correspond to the 
     width of the $\sigma$ distribution over 40 showers for each zenith angle.}
      \label{fig:meanenergy}
\end{figure}

The mean energy distribution of electrons and positrons in the specific energy range responsible for the radio signal is studied in the $({\bf v}\times {\bf B})$ direction of the shower plane. Figure~\ref{fig:meanenergy} shows this distribution at $X_{\rm max}$ for the same 4  zenith angles as previously: $\theta = 40^\circ, 60^\circ, 80^\circ, 87^\circ$ on a set of 50 simulations for each zenith angle. As expected, we observe that the most energetic particles are contained at the center of the shower while the less energetic ones are much more deflected by the Earth magnetic field. It is consistent with the expected rigidity of the particles $R = r_{\rm L}B$ which is a measure of the particle resistance to deflection by a magnetic field. It depends on the Larmor radius $r_{\rm L}$, which defines the circular motion of a particle in a uniform magnetic field, and which is given in Section~\ref{sec:scattering processes} by $r_{\rm L} \sim 3.3\times 10^4\,{\rm m}\,\sin \alpha \, (\epsilon/100\,{\rm MeV})(B/10\,\mu{\rm T})^{-1}$. The Larmor radius scales linearly with the particle energy and thus more energetic particles have greater rigidity than less energetic ones.

\subsection{Lateral extent estimation}
\begin{figure}[!ht]
\centering 
\includegraphics[width=0.7\columnwidth]{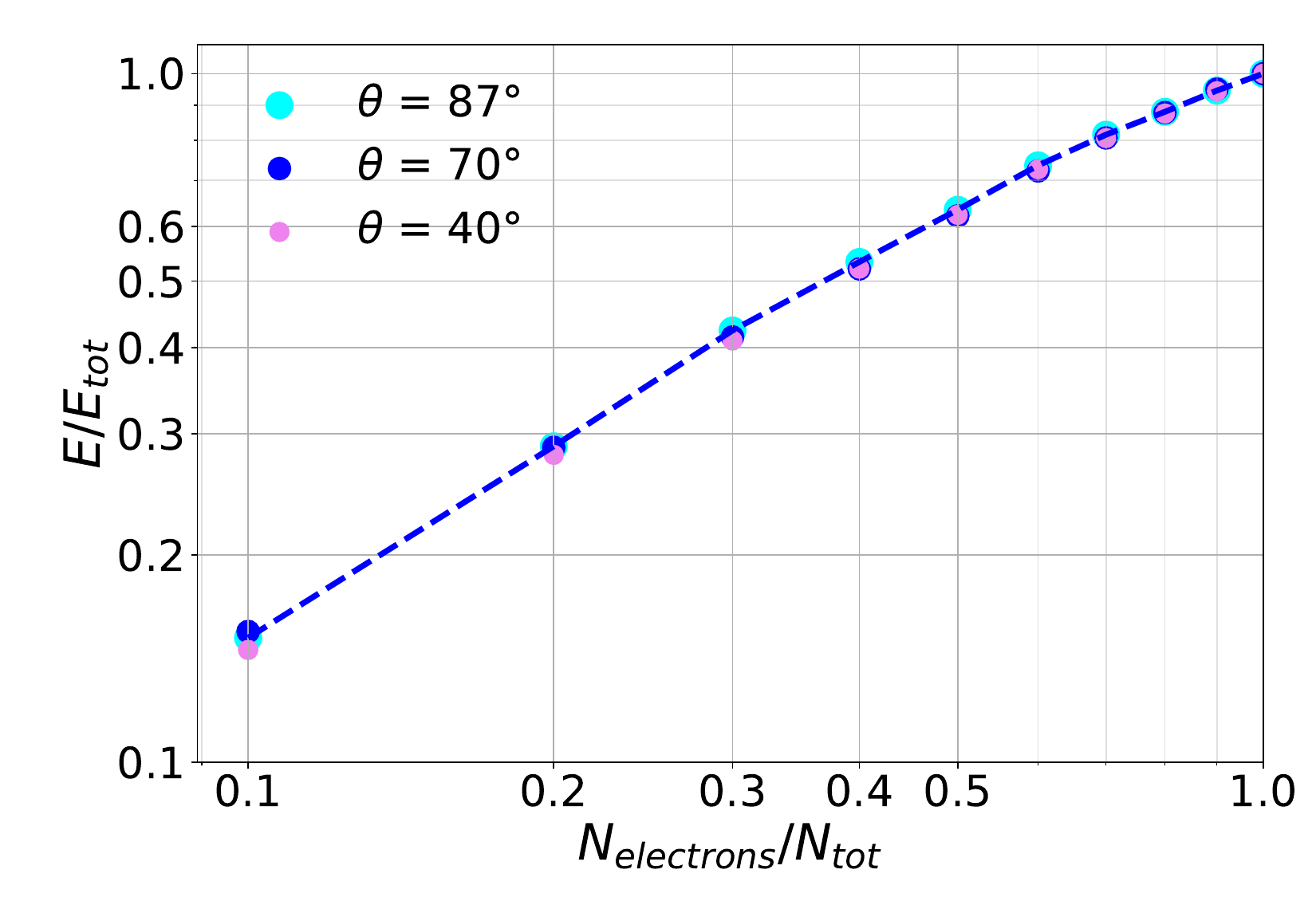}
\caption{ 
 What is the fraction of energy contained in a given population of electrons spread across the $\textbf{v} \times \textbf{B}$ axis? 
For a given density of electrons in the shower plane along the $\textbf{v} \times \textbf{B}$ axis, which corresponds to the cumulative fraction of electron population $N_{\rm electrons}/N_{\rm tot}$ along this direction, 
we represent the corresponding cumulative energy fraction $E/E_{\rm tot}$. For $\theta = 40^\circ$, only electrons in the range [10, 500] MeV are selected, for $\theta = 87^\circ$  and $\theta = 70^\circ$, the range [50, 1500] MeV is selected. 
}\label{fig:estimator}
\end{figure}

The spatial position of particles in the shower plane $[{\bf v}\times {\bf B}, {\bf v}\times ({\bf v}\times{\bf B})]$ at $X_{\rm max}$ is easily accessible with CORSIKA, as shown in Figure~\ref{fig:densityplot_twoangles}.
To estimate the lateral extension of the particles, we therefore work on the projection of the particle positions along the $\textbf{v} \times \textbf{B}$ axis, as this deflection produces a time varying current mainly responsible for the radio emission. Figure~\ref{fig:estimator} represents the cumulative fraction of electron energies from the shower core 
and the corresponding particle fractions for three different inclinations. It appears that the relationship between energy fraction and particle number at a given distance is almost a power law whatever the zenith angle.

At the distance from the shower axis where the shower contains 90\% of the particles, more than 90\% of the energy is also contained. 
This distance is thus representative of the lateral extent and is chosen as the lateral extent estimator in this work.

\begin{figure}[!ht]
\centering 
\includegraphics[width=0.7\columnwidth]{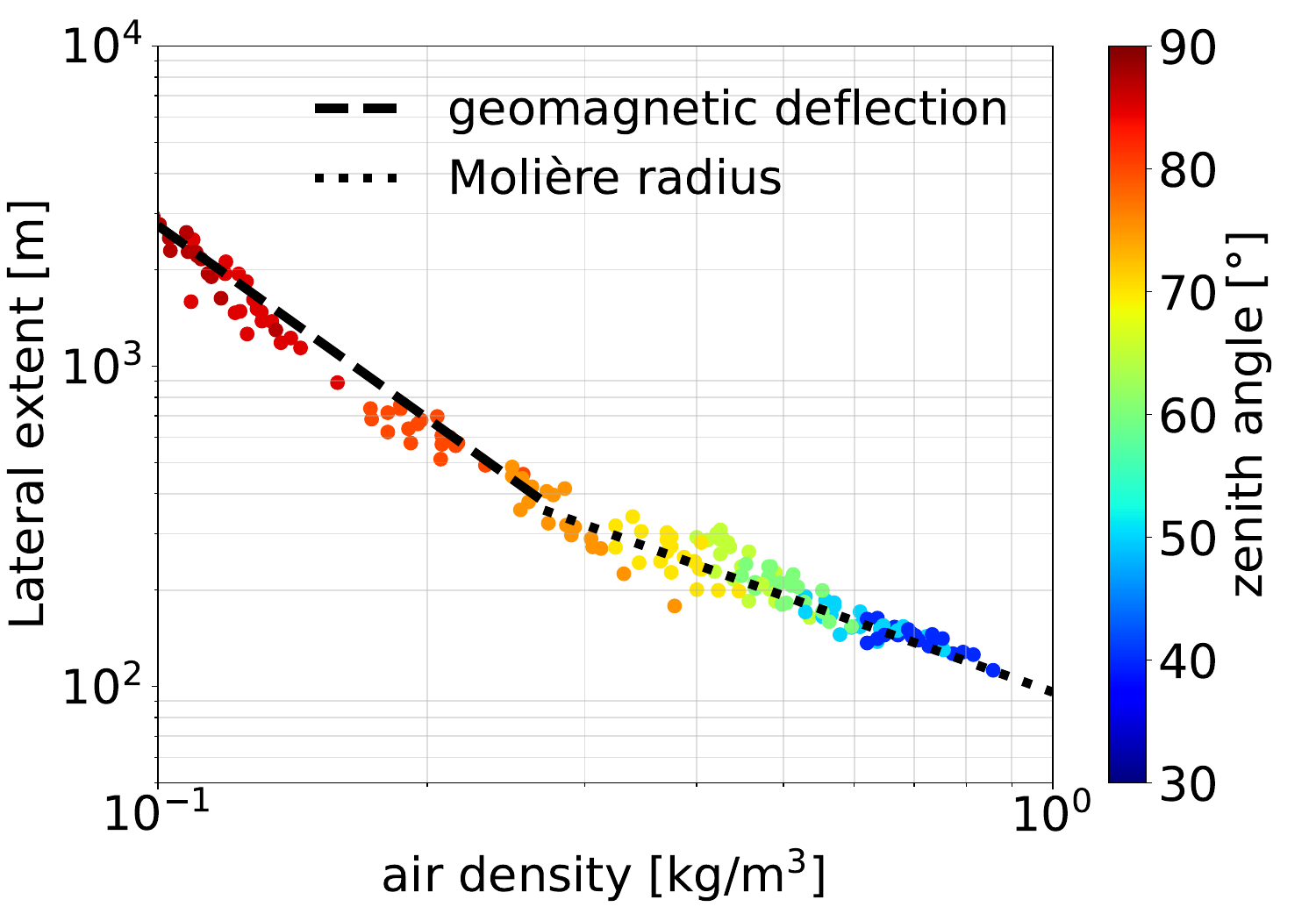}

\caption{Lateral extent of particles in the shower plane at $X_{\rm max}$ for air showers with inclinations between $30^\circ$ and $87^\circ$ on a set of 200 simulations. The air-density at $X_{\rm max}$ is also represented. The dashed line is the analytical calculation of the lateral extent given in Section~\ref{eq:lateral_extent} and the dotted line stands for the Molière radius.}\label{fig:lateral_extent}
\end{figure}

\subsection{Lateral extent as a function of inclination}

The lateral extent is computed on a set of 200 simulations obtained with CORSIKA for air showers with zenith angles varying between $30^\circ$ and $87^\circ$ --and thus for different atmospheric densities as shown in Figure~\ref{fig:lateral_extent}. According to Sections~\ref{sec:spectra} and \ref{sec:energy}, the energy of the particles responsible for the bulk of the radio emission is a function of the inclination and two different ranges of particle energies are selected: [50, 1500] MeV for particles above $70^\circ$ and [10, 500] MeV for particles below.

We observe that the lateral extent increases with zenith angle, because inclined air showers develop in thinner atmosphere than vertical ones. 
The combination of the dotted and dashed lines correspond to our analytical description given in Eq.~\ref{eq:lateral_extent_full}.
 The black dashed line corresponds to the analytical
calculation of the extension induced by the geomagnetic field, presented in Eq.~\eqref{eq:lateral_extent} for an energy of 100\,MeV. 
The corresponding particles are indeed 
representative of the 
electromagnetic content in very inclined air showers (see Section~\ref{sec:spectra}).
The dotted line is the Molière radius given in Ref.~\cite{Scholten} by $R_{\rm M} = (9.6\,\rm g/cm^{2})/\rho_{\rm air}$ which represents the spread due to scattering. 
Hence, two distinct regimes are highlighted at low and high inclinations as mentioned in Section~\ref{sec:distribution}: whereas the lateral extent of vertical air showers is driven by multiple scattering, very inclined air showers experience a drastic lateral extent increase due to the Earth's magnetic field. 

\section{Conclusion}

Using CORSIKA and CoREAS simulations and taking into account shower-to-shower fluctuations (i.e. variations on the size of the shower, on the location of $X_{\rm max}$ for air showers with exactly the same first parameters), we demonstrate the existence of two different particle energy regimes responsible for the radio signal depending on the inclination: the electric field at ground is predominantly produced by high energy particles for very inclined air showers, whereas the contribution of low energy particles increases for lower inclinations. 
We also demonstrate that the main region of radio signal emission is around $X_{\rm max}$ for all zenith angles.

 Using these results, we study the particle content of air showers subjected to a strong magnetic field $(B=56\,\mu T)$ for various inclinations in the shower plane. We find a drastic lateral extent increase for very inclined air showers in the $({\bf v}\times {\bf B})$ direction whereas this extent is constrained for vertical ones. Based on these observations, we construct a global estimator of the particle lateral extent at $X_{\rm max}$ in the two different energy regimes. The distance to the shower axis where 90\% of the considered particles are contained can be safely used as an estimator of the lateral extent. Although the energy ranges are quite approximate, they indicate a trend of the shower lateral extent,  taking into account solely the particles responsible for the radio emission.

We also propose an analytical formula of the lateral extent, valid for all air shower inclinations, which we find to be in agreement with the estimator of the lateral extent coming from our simulations. We find that the lateral extent can be modeled as a broken power-law, driven by deflections due to the Earth's magnetic field for very inclined air showers because they develop in thinner atmosphere whereas it mainly depends on diffusion due to multiple Coulomb scattering for vertical air showers. 
 
Interestingly, our results on the lateral extent transition for very inclined air showers are in line with the results of Ref.~\cite{coherence}. The study indeed reports a transition to incoherent radio emission for very inclined air showers at strong magnetic field locations, e.g., for magnetic field $B = 56\, {\rm \mu T}$, at radio wave frequency $\nu = 50\, {\rm MHz}$, for $\rho_{\rm air} < 0.27\, {\rm kg\,m^{-3}}$ ($\theta > 75^\circ$). It is argued that the drastic lateral extent increase induced for very inclined air showers subjected to a strong enough magnetic field results in a loss of coherence and in a drop of the radio signal. 
 
 This specific feature will strongly impact the reconstruction strategies of the next generation of extended radio arrays and could, for example, allow us to discriminate between cosmic-ray air showers and neutrinos that also have very inclined trajectories but which develop in much denser atmosphere \cite{coherence}.

\acknowledgments

The authors thank the anonymous referee, T. Huege, V. Niess and M. Tueros for their careful reading and the fruitful discussions, as well as the GRAND-Paris team, K. de Vries, V. Decoene, and the GRAND team at KIT. This work was supported by the Programme National des Hautes Energies
of CNRS/INSU with INP and IN2P3, co-funded by CEA and CNES. Simulations were performed using the computing resources at the CCIN2P3 Computing Centre (Lyon/Villeurbanne – France), partnership between CNRS/IN2P3 and CEA/DSM/Irfu. This work was supported by the PHC TOURNESOL program 48705Z and is part of the NuTRIG project, supported by the Agence Nationale de la Recherche (ANR-21-CE31-0025; France) and the Deutsche Forschungsgemeinschaft (DFG; Projektnummer 490843803; Germany).

\appendix
\section{Effects of frequency range}
\label{sec:frequency}

Throughout this study, our results have been derived taking into account signals at all frequencies. However, radio air shower experiments are only be sensitive to the radio emission in a specific frequency band. We examine in this section the effect of filtering the radio emission to a broad band which is typically available for radio experiments.

\begin{figure}[!t]
     \centering
         \includegraphics[height=0.5\columnwidth]{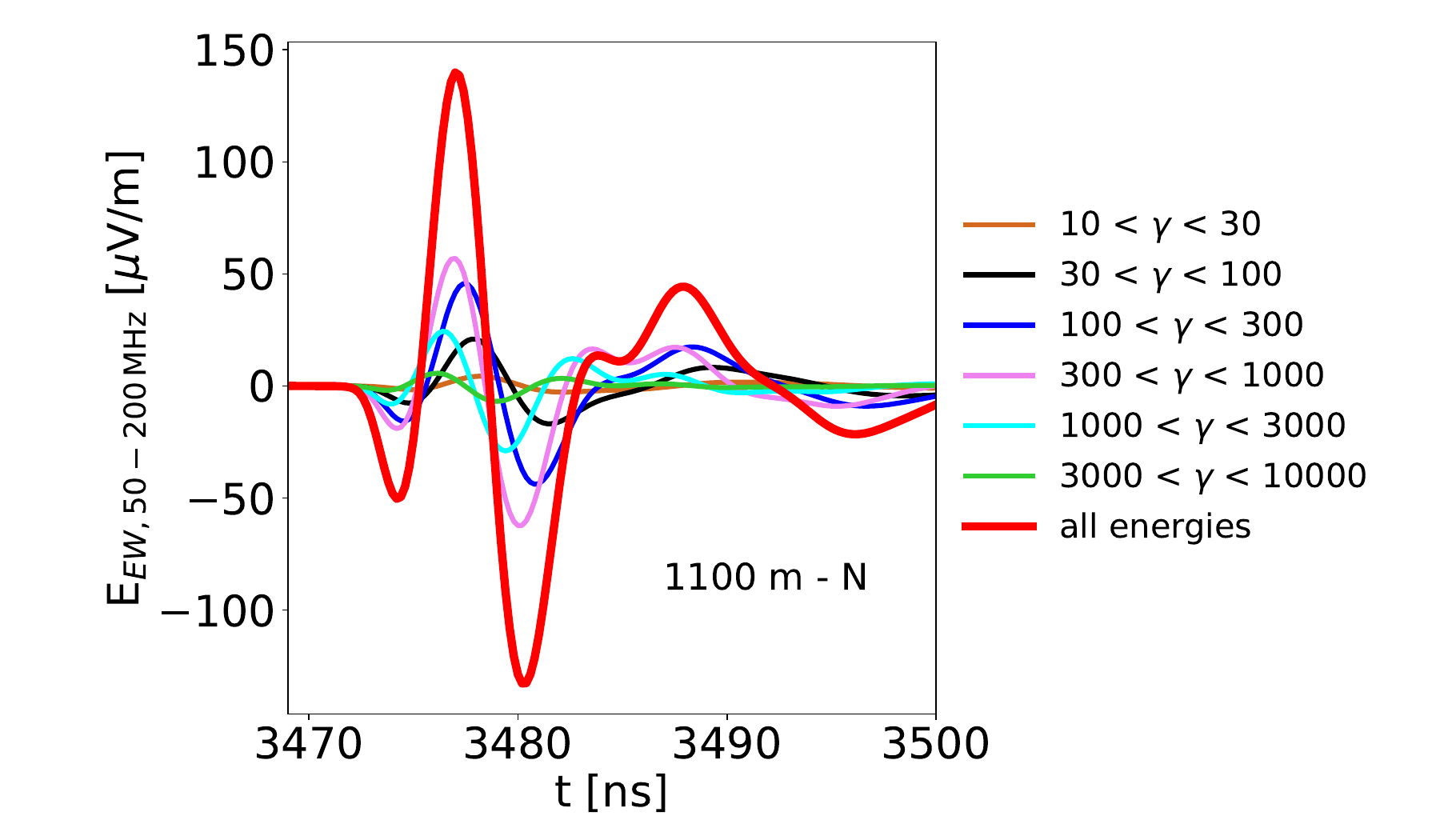}            
      \captionof{figure}{Traces of the East/West polarisation of the radio signal filtered in the [50 - 200]\,MHz frequency range and created at different electron/positron energy (Lorentz factor) ranges for an air shower with inclination $\theta=70^\circ$ and azimuth $\phi=0^\circ$. The primary particle is a proton with energy $10^{17}\,\rm eV$. The antenna is located at a distance of 1100\,m of the shower core on the North axis, where the radio signal is maximum. The lines represent the electric fields produced by particles in specific energy ranges as indicated.}
    \label{fig:trace70_filtered}
\end{figure}

We use the same set of simulated data as previously, as described in Section~\ref{section:sims}, and we model the frequency response of the antenna by using the passband Butterworth filter \cite{Butterworth1930} at order 5, in the [50, 200] MHz frequency range. The time trace of Figure~\ref{fig:trace70}, filtered in this range, is shown in Figure~\ref{fig:trace70_filtered}.

\begin{figure}[!b]
     \centering
     \includegraphics[height=0.5\columnwidth]{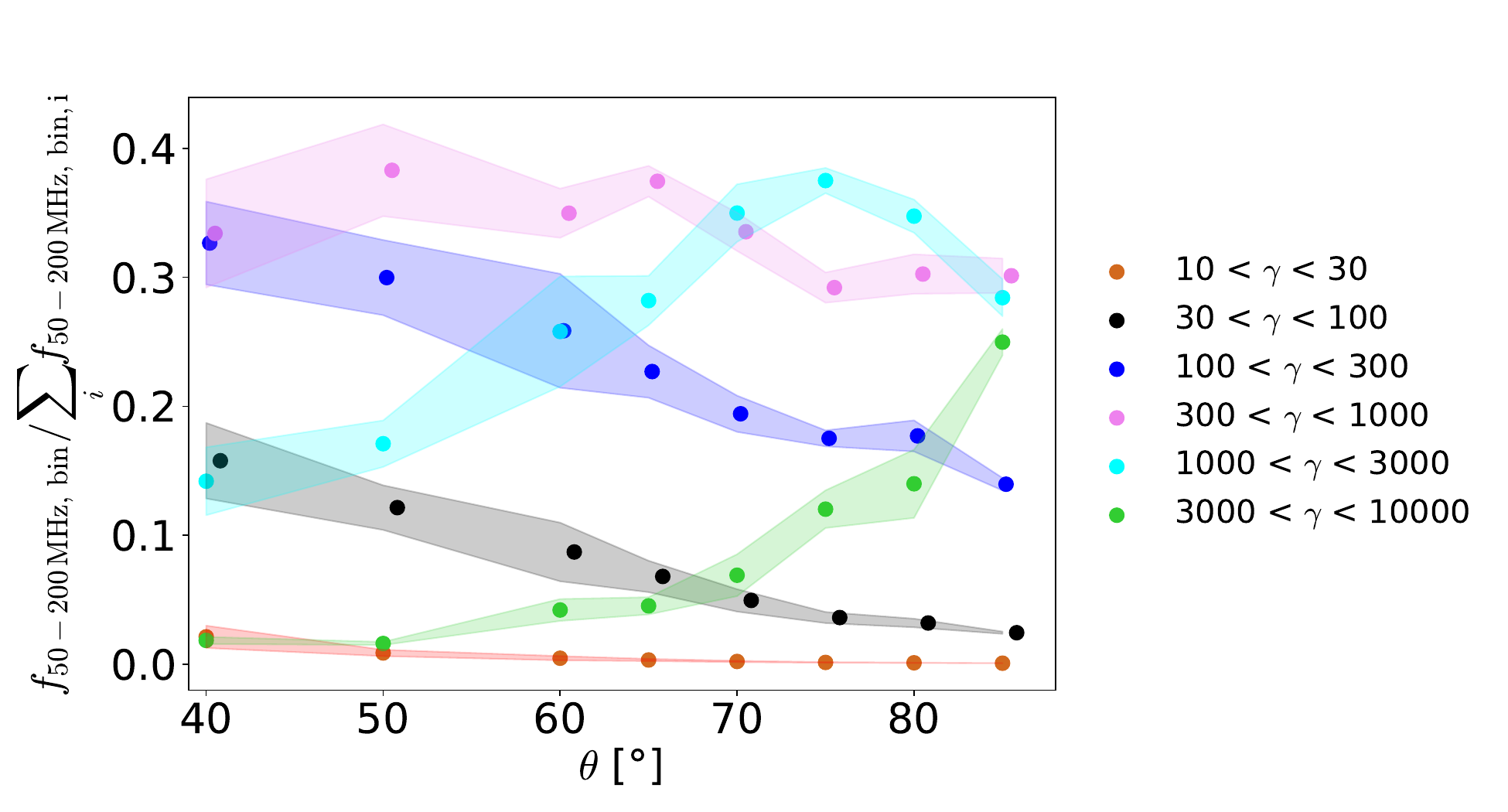}
     \includegraphics[height=0.5\columnwidth]{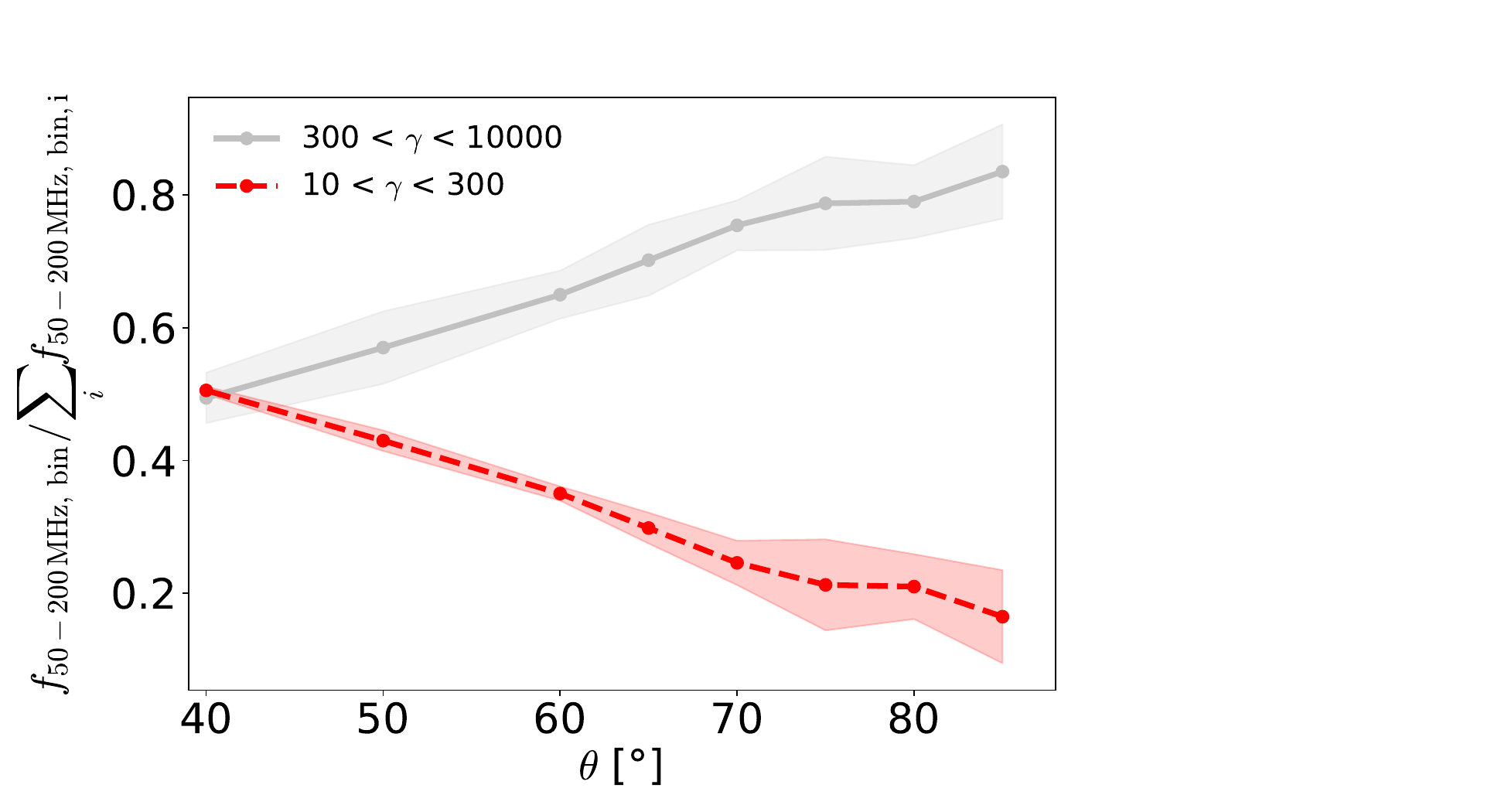}
     \captionof{figure}{{\it (Top)} Contribution of different particle energy regimes responsible for the radio signal in the  [50, 200] MHz frequency range. Each point corresponds to the mean value of the normalized energy fluence in a given particle energy range on a set of 40 simulations and the shaded regions correspond to the corresponding 
     width of the distribution. {\it (Bottom)} Sum of the contribution of particle energy regimes for $ 10 < \gamma <300 $ and $ 300 < \gamma < 10000$, with corresponding shaded error bar region.}
      \label{fig:energy_filtered}
\end{figure}

By applying the same method as in Section~\ref{sec:energy}, the energy fluence received by the antenna in each energy pulse is computed. Figure~\ref{fig:energy_filtered} shows the contribution to the East/West polarisation of the electric field for particles in different energy bins. Results similar to the raw data, i.e. to the unfiltered signal, are found. 

\section{Effects of interactions between particles}
\label{sec:interferences}
\begin{figure}[!t]
     \centering
     \includegraphics[height=0.5\columnwidth]{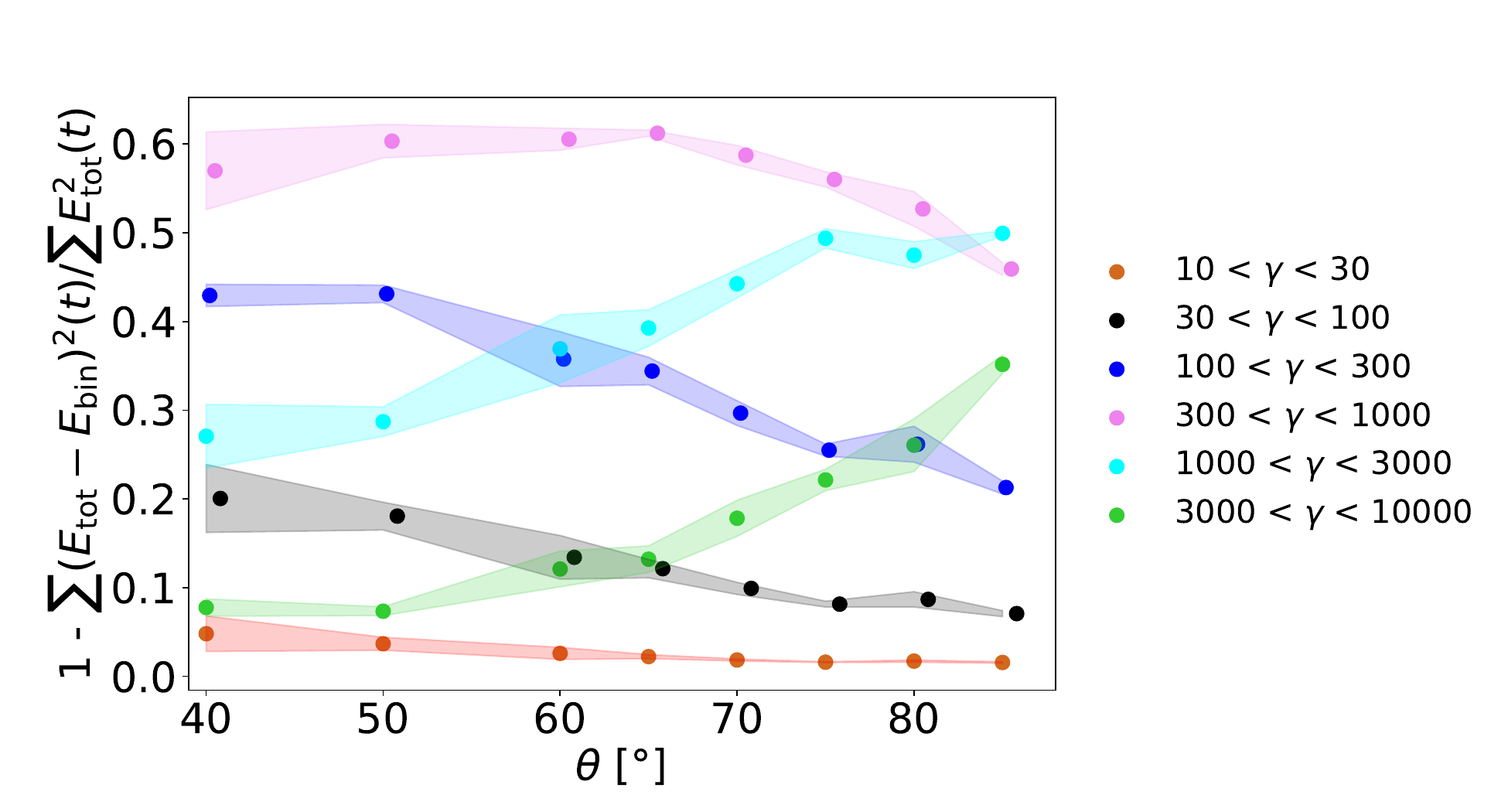}
     \captionof{figure}{Contribution of different particle energy regimes responsible for the radio signal taking into account the possible destructive interferences between particles of different energy ranges. Each point corresponds to the mean value of the normalized energy fluence in a given particle energy range on a set of 40 simulations and the shaded regions correspond to the corresponding 
     width of the distribution.}
      \label{fig:energy_interferences}
\end{figure}

By studying the energy fluence of each radio pulse independently as in Section~\ref{sec:energy}, we may introduce a bias by ignoring destructive interferences between particles of different energy ranges. We thus check the influence of this effect by using an alternative method to estimate the contribution of each energy bin in the radio emission at ground. We have access to the radio signal created by all the particles of the air shower and we can compute the total energy fluence $f_{\rm tot} \propto \sum_{i} E_{\rm tot}^{2}(t_{i}) \propto  \sum_{i} (E_{\rm EW, tot}^{2}(t_{i}) + E_{\rm NS, tot}^{2}(t_{i}) + E_{\rm vertical, tot}^{2}(t_{i})) $ where all the collective effects are intrisically taken into account. The contribution of particles in a specific bin is estimated by computing $ 1 - \sum_{i} (E_{\rm tot} - E_{\rm bin})^{2}(t_{i})/\sum_{i} E_{\rm tot}^{2}(t_{i})$. 
The contribution to the total radio emission of the  different pulses are shown in Fig.~\ref{fig:energy_interferences}. The closer the ratio is to one, the more the particles of a bin participate to the radio signal. These results are similar to those obtained by studying the energy bins independently.

\clearpage
\bibliographystyle{JHEP}
\bibliography{biblio.bib}

\providecommand{\href}[2]{#2}\begingroup\raggedright\begin{thebibliography}{10}

\bibitem{AERA}
T.~Huege, \emph{Radio detection of cosmic ray air showers in the digital era}, \href{https://doi.org/10.1016/j.physrep.2016.02.001}{\emph{Physics Reports} {\bfseries 620} (2016) 1}.

\bibitem{Schr_der_2017}
F.G.~Schröder, \emph{Radio detection of cosmic-ray air showers and high-energy neutrinos}, \href{https://doi.org/10.1016/j.ppnp.2016.12.002}{\emph{Progress in Particle and Nuclear Physics} {\bfseries 93} (2017) 1}.

\bibitem{Nelles_2015}
A.~Nelles, S.~Buitink, A.~Corstanje, J.~Enriquez, H.~Falcke, J.~Hörandel et~al., \emph{The radio emission pattern of air showers as measured with {LOFAR}{\textemdash}a tool for the reconstruction of the energy and the shower maximum}, \href{https://doi.org/10.1088/1475-7516/2015/05/018}{\emph{Journal of Cosmology and Astroparticle Physics} {\bfseries 2015} (2015) 018}.

\bibitem{GRAND_paper}
J.~{\'{A}}lvarez-Mu{\~{n}}iz, R.A.~Batista, A.B.~V., J.~Bolmont, M.~Bustamante, W.~Carvalho et~al., \emph{The giant radio array for neutrino detection ({GRAND}): Science and design}, \href{https://doi.org/10.1007/s11433-018-9385-7}{\emph{Science China Physics, Mechanics and Astronomy} {\bfseries 63} (2019) }.

\bibitem{beacon}
D.~Southall, C.~Deaconu, V.~Decoene, E.~Oberla, A.~Zeolla, J.~Alvarez-Mu{\~{n}}iz et~al., \emph{Design and initial performance of the prototype for the {BEACON} instrument for detection of ultrahigh energy particles}, \href{https://doi.org/10.1016/j.nima.2022.167889}{\emph{Nuclear Instruments and Methods in Physics Research Section A: Accelerators, Spectrometers, Detectors and Associated Equipment} {\bfseries 1048} (2023) 167889}.

\bibitem{AugerPrime}
A.~{Castellina} and {Pierre Auger Collaboration}, \emph{{AugerPrime: the Pierre Auger Observatory Upgrade}},  in \emph{European Physical Journal Web of Conferences}, vol.~210 of \emph{European Physical Journal Web of Conferences}, p.~06002, Oct., 2019, \href{https://doi.org/10.1051/epjconf/201921006002}{DOI} [\href{https://arxiv.org/abs/1905.04472}{{\ttfamily 1905.04472}}].

\bibitem{Chiche}
S.~Chiche et~al., \emph{{New paradigm and radio signatures for very inclined air showers}},  in \emph{{ICRC} Conference Proceedings}, 2023.

\bibitem{Scholten}
O.~Scholten, K.~Werner and F.~Rusydi, \emph{A macroscopic description of coherent geo-magnetic radiation from cosmic-ray air showers}, \href{https://doi.org/10.1016/j.astropartphys.2007.11.012}{\emph{Astroparticle Physics} {\bfseries 29} (2008) 94}.

\bibitem{Schroder}
F.G.~Schröder, \emph{Radio detection of cosmic-ray air showers and high-energy neutrinos}, \href{https://doi.org/10.1016/j.ppnp.2016.12.002}{\emph{Progress in Particle and Nuclear Physics} {\bfseries 93} (2017) 1}.

\bibitem{Engel}
T.K.~{Gaisser}, R.~{Engel} and E.~{Resconi}, \emph{{Cosmic Rays and Particle Physics}} (2016).

\bibitem{Hillas}
A.M.~{Hillas}, \emph{{The Origin of Ultra-High-Energy Cosmic Rays}}, \href{https://doi.org/10.1146/annurev.aa.22.090184.002233}{\emph{Annual Review of Astron and Astrophys} {\bfseries 22} (1984) 425}.

\bibitem{CORSIKA}
D.~{Heck}, J.~{Knapp}, J.N.~{Capdevielle}, G.~{Schatz} and T.~{Thouw}, \emph{{CORSIKA: a Monte Carlo code to simulate extensive air showers.}} (1998).

\bibitem{Joao}
J.~Torres~de Mello~Neto, \emph{{The Giant Array for Neutrino Detection (GRAND)}},  in \emph{{ICRC} Conference Proceedings}, 2023.

\bibitem{CoREAS}
T.~Huege, M.~Ludwig and C.W.~James, \emph{Simulating radio emission from air showers with {CoREAS}},  in \emph{{AIP} Conference Proceedings}, {AIP}, 2013, \href{https://doi.org/10.1063/1.4807534}{DOI}.

\bibitem{Huege_2007}
T.~Huege, R.~Ulrich and R.~Engel, \emph{Monte carlo simulations of geosynchrotron radio emission from corsika-simulated air showers}, \href{https://doi.org/10.1016/j.astropartphys.2007.01.006}{\emph{Astroparticle Physics} {\bfseries 27} (2007) 392–405}.

\bibitem{2011PhDT156L}
M.~{Ludwig}, \emph{{Modelling of radio emission from cosmic ray air showers}}, Ph.D. thesis, Karlsruhe Institute of Technology, Germany, June, 2011.

\bibitem{Glaser}
C.~Glaser, M.~Erdmann, J.R.~Hörandel, T.~Huege and J.~Schulz, \emph{Simulation of radiation energy release in air showers}, \href{https://doi.org/10.1088/1475-7516/2016/09/024}{\emph{Journal of Cosmology and Astroparticle Physics} {\bfseries 2016} (2016) 024}.

\bibitem{energy_radiation}
A.~Aab, P.~Abreu, M.~Aglietta, E.~Ahn, I.A.~Samarai, I.~Albuquerque et~al., \emph{Measurement of the radiation energy in the radio signal of extensive air showers as a universal estimator of cosmic-ray energy}, \href{https://doi.org/10.1103/physrevlett.116.241101}{\emph{Physical Review Letters} {\bfseries 116} (2016) }.

\bibitem{Deconene2020}
V.~{Decoene}, \emph{{Sources and detection of high-energy cosmic events}}, Ph.D. thesis, Sorbonne Université, France, 2020.

\bibitem{PhysRevD.98.030001}
{\scshape Particle Data Group} collaboration, \emph{Review of particle physics}, \href{https://doi.org/10.1103/PhysRevD.98.030001}{\emph{Phys. Rev. D} {\bfseries 98} (2018) 030001}.

\bibitem{coherence}
S.~Chiche et~al., \emph{Loss of coherence and change in emission physics for radio emission from very inclined air showers}, {\emph{Physical Review Letters (\textit{submitted})} }.

\bibitem{Butterworth1930}
S.~Butterworth, \emph{On the {{Theory}} of {{Filter Amplifiers}}}, {\emph{Experimental Wireless \& the Wireless Engineer} {\bfseries 7} (1930) 536}.

\end{thebibliography}\endgroup






\end{document}